\documentclass[superscriptaddress,aps,prd,twocolumn,showpacs,nofootinbib,floatfix]{revtex4-2}
\usepackage{float}
\usepackage{multirow}
\usepackage{amsmath}
\usepackage{amssymb}
\usepackage{mathtools}
\usepackage{graphicx}
\usepackage{xcolor}
\usepackage{stackrel}
\usepackage{natbib}
\definecolor{tl}{RGB}{0,180,120}
\usepackage[colorlinks,citecolor=tl,linkcolor=red,urlcolor=magenta,unicode=true]{hyperref}

\begin{document}
\title{Analyzing the 21-cm signal brightness temperature in the Universe with inhomogeneities}
\author{Shashank Shekhar Pandey}
\email{shashankpandey7347@gmail.com}
\affiliation{Department of Astrophysics and High Energy Physics, 
S. N. Bose National Centre for Basic Sciences, Block - JD, Sector - III, 
Salt Lake, Kolkata-700106, India}
\author{Ashadul Halder}   \email{ashadul.halder@gmail.com}
\affiliation{Department of Astrophysics and High Energy Physics, 
S. N. Bose National Centre for Basic Sciences, Block - JD, Sector - III, 
Salt Lake, Kolkata-700106, India}
\author{A. S. Majumdar}   \email{archan@bose.res.in}
\affiliation{Department of Astrophysics and High Energy Physics, 
S. N. Bose National Centre for Basic Sciences, Block - JD, Sector - III, 
Salt Lake, Kolkata-700106, India}
\date{\today}
\begin{abstract}
\begin{center}
\textbf{Abstract}
\end{center}
We explore the 21-cm signal in our Universe containing inhomogeneous matter distribution at considerably large scales. Employing Buchert's averaging procedure in the context of a model of spacetime with multiple inhomogeneous domains, we evaluate the effect of our model parameters on the observable 21-cm signal brightness temperature. Our model parameters are constrained through the  Markov Chain Monte Carlo method using the Union 2.1 supernova Ia observational data. We find that a significant dip in the brightness temperature compared to the $\Lambda$CDM prediction could arise as an effect of the inhomogeneities present in the Universe.

\end{abstract}

\maketitle
\section{Introduction }\label{sec:intro}

21 cm cosmology has been an essential tool in studying the physics of the cosmic dark age for quite some time now \cite{FURLANETTO2006181, morales_review, Pritchard_2012}. It is a unique probe of the re-ionization epoch of the Universe \cite{Purcell1951, OORT1951}. 21-cm is the wavelength corresponding to the energy shift due to hyperfine splitting in the ground state of neutral Hydrogen, the most abundant element in the Universe. It occupies $\sim 75\%$ of the entire baryonic allocation of the Universe. The transition between hydrogen atoms' electronic spin states (s = 0, 1) generates the 21-cm ($\sim$ 1.42 GHz) hyperfine spectrum.

The brightness temperature $T_{21}$ associated with this spectrum is a function of $T_s - T_{\gamma}$ where $T_{\gamma}$ is the cosmic microwave background temperature given by $ T_{\gamma} = 2.725(1+z)K$ and $T_s$ is the spin temperature \cite{Pritchard_2012, FURLANETTO2006181, morales_review}. Recently, the ``Experiment to Detect the Global Epoch of Re-ionization Signature" (EDGES) \cite{EDGES_Bowman2018} generated quite a bit of excitement in the field with its reported $T_{21}$ in the redshift range $14<z<20$  to be $-500^{+200}_{-500}$ mK. However, the subsequent SARAS experiment \cite{SARAS_Singh_2018} failed to detect the EDGES 21 cm signal \cite{SARAS_Singh2022}. 

The $\Lambda$CDM model of standard cosmology estimates a brightness temperature of about $\approx -200$ mK without any additional thermal contribution. To account for any additional cooling,  various phenomenological effects in the early Universe have been employed \cite{Ashadul_mnras, Ashadul_PRD, Halder_2021, Clark_2018, Datta_PRD, yang2020}, including exotic models of dark matter \cite{munoz_haimound, Halder_2021, Datta_PRD} and dark energy \cite{Halder_2022}. The 21-cm signal provides an avenue towards discerning several physical phenomena of this epoch, such as evaporating primordial black holes (PBHs) \cite{Clark_2018, yang2020, Ashadul_mnras, Akash_PBH_IISc}, baryon-dark matter scattering \cite{munoz_haimound, Halder_2021, Datta_PRD}, and neutrino physics \cite{neutrino}.

Although the $\Lambda$CDM model has been highly successful in establishing the basic tenets of standard cosmology, it has been confronted with certain discordant observations in recent times, notably among them the so-called Hubble tension \cite{Riess_2021, Freedman_2021}. Moreover, recent observations of large-scale structures (LSS) indicate the scope of including additional features within the framework of standard cosmology. Although the Universe may be uniform and isotropic at the largest scales, various astrophysical surveys have revealed prominent matter distribution inhomogeneities up to slightly smaller scales \cite{Labini_2009}. Significant (more than 3$\sigma$) deviations from the $\Lambda$CDM mock catalogues have been reported \cite{wiegand_scale} on luminous red galaxy samples as large as 500 $h^{-1}$ Mpc. Recently,  a giant arc of galaxies spanning $\sim$ 1 Gpc (proper size, present epoch) has been observed \cite{lopez}. 

Such observed deviations in LSS from the assumed smooth homogeneous $\Lambda$CDM paradigm may necessitate the inclusion of the impact of inhomogeneities in the analyses of cosmological phenomena. An averaging procedure is required to incorporate the effect of inhomogeneities in the analysis. Various averaging schemes have been proposed in the literature \cite{Ellis1984, Futamase, Zalaletdinov1992, Zalaletdinov1993, Gasperini_2011}. Buchert introduced a simplified averaging procedure \cite{Buchert, Buchert2001} restricted to scalar quantities on space-like hypersurfaces. Several studies employing the Buchert averaging procedure have been done to investigate the effect of backreaction of inhomogeneities on cosmological dynamics, including attempts to explain the current accelerated expansion of the Universe without invoking exotic physics \cite{Coley, Buchert, Buchert2001, Korzynski_2010, Clifton, Skarke, Buchert_2015, Buchert4, Weigand_et_al, Rasanen_2004, wiltshire, Kolb_2006, Koksbang_2019, Koksbang2, Koksbang_PRL, Rasanen_2008, bose, Bose2013, Ali_2017, Pandey_2022, Pandey2023, Koksbang5, Koksbang6, Koksbang7, Koksbang8, Halder_2023, Koksbang2023}. 

Buchert's averaging procedure for evaluating the effect of backreaction of matter distribution inhomogeneities offers the prospect of relating theoretically calculated spatially averaged quantities to observationally measurable quantities, such as redshift-distance relations \cite{rasanen1, rasanen2, Koksbang_2019, Koksbang2, Koksbang3, Koksbang4}. Analyses of light wave propagation in the presence of inhomogeneities have revealed interesting features in the modified redshift-distance relation due to the backreaction effect \cite{Koksbang_2019, Koksbang2, Koksbang3}. Similar effects have also been derived in the context of gravitational waves emitted from compact binaries propagating through the inhomogeneous matter distribution \cite{Pandey_2022, Pandey2023}. In the present work, we are thus motivated to explore the 21-cm signal under the impact of backreaction from matter inhomogeneities employing the Buchert averaging framework.

Specifically, in the present analysis, we employ a model of spacetime with a spectrum of matter distribution inhomogeneities in multiple domains, which can imitate our actual Universe more realistically compared to earlier analyses of cosmological dynamics under backreaction, which have primarily relied on toy two-domain models \cite{bose, Bose2013, Ali_2017, Pandey_2022, Koksbang_2019}. We aim to theoretically analyze the 21-cm  signal in the context of modified Hubble dynamics due to the effect of backreaction from matter distribution inhomogeneities evaluated using the Buchert formalism. We constrain our model parameters by performing Markov Chain Monte Carlo analysis using the Union 2.1 supernova Ia data to determine the model parameters' best fit and optimum values. We compute the 21-cm brightness temperature, which reveals a significant dip compared to the $\Lambda$CDM prediction in the redshift range $14<z<20$, without invoking any additional non-standard cosmological effects or exotic physics. 

The paper is organized as follows. We briefly introduce the formalism for evaluating the 21-cm signal brightness temperature in (\autoref{sec:21cm}). Next, we briefly outline Buchert's averaging procedure in (\autoref{sec:buchert}). Our model of multidomain inhomogeneities is presented, leading to modified Hubble dynamics in (\autoref{sec:our model}). We then present our analysis of the 21-cm signal in the context of our multidomain model in (\autoref{sec:analysis}). In (\autoref{sec:obs}), we use the Union 2.1 supernova Ia data to constrain our model parameters and compute the 21 cm brightness temperature in the redshift range $14<z<20$ using the best fit and optimum values of our model parameters. Finally, we summarize our results in (\autoref{sec:results}).

\section{\label{sec:21cm} 21-cm brightness temperature}

The 21-cm absorption line of the hydrogen atom is generated by the transition of an electron between the two hyperfine spin states (spin 0 and spin 1). This 21-cm line has a characteristic temperature associated with it called the brightness temperature, $T_{21}$, which represents the intensity of the 21-cm line as a function of the cosmological redshift $z$. The expression for $T_{21}$ is given by \cite{Ashadul_PRD, munoz_21cm},
\begin{equation}\label{eq:T21}
    T_{21} = \frac{T_s - T_\gamma}{1+z}(1-e^{-\tau(z)}),
\end{equation}
where $T_s$ is the 21-cm spin temperature at redshift $z$, $T_\gamma$ is the cosmic microwave background (CMB) temperature $(T_\gamma = 2.725(1+z)K)$ and $\tau(z)$ is the optical depth of the inter-galactic medium (IGM). $\tau(z)$ is given by \cite{munoz_21cm},
\begin{equation}\label{eq:tau}
    \tau(z) = \frac{3}{32\pi}\frac{T_*}{T_s}n_{HI}\lambda_{21}^3\frac{A_{10}}{H(z)+(1+z)\delta_rv_r}
\end{equation}
where $T_* = hc/(k_B\lambda_{21}) = 0.068$~K, $A_{10} = 2.85\times 10^{-15} s^{-1}$ is the Einstein coefficient \cite{Haimoud_Hirata_2010}, $\lambda_{21} \approx 21$ cm, $n_{HI}$ is the local neutral hydrogen density, $H(z)$ is the Hubble parameter and $\delta_rv_r$ is the radial gradient of the peculiar velocity. The above equations are derived from the principles of atomic physics and radiative transfer. The expression for optical depth has a term for the gradient of the proper velocity along the line of sight, and this term includes both the Hubble expansion and the peculiar velocity, as can be seen in (\autoref{eq:tau}) \cite{FURLANETTO2006181}. This is how $H(z)$, the Hubble parameter, which measures the Universe's rate of expansion, enters the analysis. Since $H(z)$ itself depends on the model of cosmology, the values of the quantities
of interest turn out to be different for different models of cosmology.

The spin temperature $T_s$ is related to the ratio of the number density of hydrogen atoms in excited and ground states and is given by \cite{munoz_21cm},
\begin{equation}\label{eq:Ts1}
    \frac{n_1}{n_0} = 3 e^{-T_*/T_s}
\end{equation}
where $n_1$ and $n_0$ are the number densities of hydrogen atoms in excited and ground states, respectively. In equilibrium, $T_s$ is given by \cite{Zaldarriaga_2004, Clark_2018},

\begin{equation}\label{eq:Ts2}
    T_s = \frac{T_\gamma + y_c T_b + y_{Ly\alpha}T_{Ly\alpha}}{1+y_c+y_{Ly\alpha}}
\end{equation}
where $y_c$ is the collisional coupling parameter, $T_b$ is the baryon temperature, $y_{Ly\alpha}$ represents the Wouthuysen-Field effect \cite{Wouthuysen, Field} and $T_{Ly\alpha}$ is the Lyman-$\alpha$ ($Ly\alpha$) background temperature. The coefficients $y_c$ and $y_{Ly\alpha}$ are given by $y_c = \frac{C_{10}T_*}{A_{10}T_b}$ and $y_{Ly\alpha} = \frac{P_{10}T_*}{A_{10}T_{Ly\alpha}}$ \cite{Kuhlen_2006}. Here, $C_{10}$ is the collisional de-excitation rate of the triplet hyperfine level, $P_{10}\approx1.3\times10^{-12}S_\alpha J_{-21}s^{-1}$ is the indirect de-excitation rate due to Ly-$\alpha$ absorption, $S_\alpha$ is a factor of order unity that incorporates spectral distortions \cite{Hirata_2006} and $J_{-21}$ is the Lyman-$\alpha$ background intensity written in units of $10^{-21}$ erg $\rm{cm^{-2}}$ $\rm{s^{-1}}$ $\rm{Hz^{-1}}$ $\rm{sr^{-1}}$, and can be estimated by the procedure mentioned in \cite{Clark_2018, Ciardi_2003}. 

The baryon temperature $T_b$ can be obtained from the standard evolution equations of $T_b$ and $x_e$, the ionization fraction. The ionization fraction $x_e$ is given by $x_e = n_e/n_H$, where $n_e$ and $n_H$ are the number densities of free electron and Hydrogen, respectively, is an important quantity in estimating thermal evolution. Here, we will not consider exotic physics or non-standard processes like dark matter decay. We just consider standard evolution equations. These equations governing thermal evolution are given by \cite{Clark_2018, haimoud_hirata_2011, madhava_21cm},
\begin{equation}
    (1+z)\frac{dT_b}{dz} = 2T_b + \frac{\Gamma_c}{H(z)}(T_b - T_\gamma), \label{eq:Tb}
\end{equation}

\begin{equation}
    (1+z)\frac{dx_e}{dz} = \frac{C_P}{H(z)}\left(n_H\alpha_Bx_e^2-4(1-x_e)\beta_Be^{-\frac{3E_0}{4k_BT_\gamma}}\right), \label{eq:xe}
\end{equation}
where $\Gamma_c$ $\left(\Gamma_c = \frac{8\sigma_Ta_rT^4_\gamma x_e}{3(1+f_{He}+x_e)m_ec}\right)$ describes the Compton interaction rate ($\sigma_T$ is the Thomson scattering cross-section, $a_r$ is the radiation constant, $f_{He}$ is the fractional abundance of Helium, $m_e$ is the mass of an electron), $C_P$ is the Peebles $C$ factor \cite{Peebles, haimoud_hirata_2011}, $E_0 = 13.6$ eV, $k_B$ is the Boltzmann constant and $\alpha_B$ and $\beta_B$ are the  recombination and ionization coefficients, respectively. The Peebles C factor is given by \cite{haimoud_hirata_2011},
\begin{equation}\label{eq:C_P}
    C_P = \frac{\frac{3}{4}R_{Ly\alpha} + \frac{1}{4}\Lambda_{2s1s}}{\beta_B+\frac{3}{4}R_{Ly\alpha} + \frac{1}{4}\Lambda_{2s1s}}
\end{equation}
where $R_{Ly\alpha}$ represents the rate of escape of Ly $\alpha$ photons, $R_{Ly\alpha} = 8\pi H/\left(3n_H(1-x_e)\lambda^3_{Ly\alpha}\right)$, $n_H$ is the total number density of Hydrogen and $\Lambda_{2s,1s}\approx 8.22 \rm{s}^{-1}$ \cite{haimoud_hirata_2011}. $\alpha_B$ and $\beta_B$ can be calculated using the procedure mentioned in \cite{Ashadul_mnras, Ashadul_PRD}. Similar to the case of (\autoref{eq:T21}) and (\autoref{eq:tau}), the cosmological dependence of the equations (\autoref{eq:Tb} and \autoref{eq:xe})  are enshrined in $H(z)$, the Hubble parameter.

\section{Buchert's backreaction formalism}\label{sec:buchert}

Buchert's averaging procedure that we are using here is for the pressureless dust universe model \cite{Buchert, buchert_rasanen}. Buchert's backreaction formalism simplifies the averaging problem by considering only scalar quantities to average. The spacetime is divided into flow-orthogonal hypersurfaces with the line element \cite{Buchert, Weigand_et_al}
\begin{equation}\label{eq:line_element}
    ds^2 = -dt^2+g_{ij}dX^idX^j,
\end{equation}
where $t$ is the proper time, $X^i$ are Gaussian normal coordinates in the hypersurfaces and $g_{ij}$ is the spatial three-metric of the hypersurfaces of constant $t$. The volume of a compact spatial domain $\mathcal{D}$ on these hypersurfaces is defined as,
\begin{equation}\label{eq:volD}
    |\mathcal{D}|_g := \int_{\mathcal{D}} d\mu_g
\end{equation}
where $d\mu_g:= \sqrt{\prescript{(3)}{}{g(t,X^1,X^2,X^3)}}dX^1dX^2dX^3$. Now, we define a dimensionless (`effective') scale factor
\begin{equation}\label{eq:scale_factor}
    a_{\mathcal{D}}(t) := \left(\frac{|\mathcal{D}|_g}{|\mathcal{D}_i|_g}\right)^{1/3},
\end{equation}
normalized by the volume of the initial domain $|\mathcal{D}_i|_g$, which can be considered the domain's volume at the present time, $t_0$. The average over a scalar quantity $f$ is defined as
\begin{equation}\label{eq:average_scalar}
    \langle f\rangle_{\mathcal{D}}(t) := \frac{\int_{\mathcal{D}}f(t,X^1,X^2,X^3)d\mu_g}{\int_{\mathcal{D}}d\mu_g}
\end{equation}

Using this averaging procedure and Einstein equations along with the continuity equation, Hamiltonian constraint and Raychaudhuri equation,  give us evolution equations,
\begin{eqnarray}\label{eq:aD_ddot}
    3\frac{\ddot{a_{\mathcal{D}}}}{a_{\mathcal{D}}} = -4\pi G\langle\rho\rangle_{\mathcal{D}} + \mathcal{Q}_{\mathcal{D}} \\
    3H^2_{\mathcal{D}} = 8\pi G\langle\rho\rangle_{\mathcal{D}} - \frac{1}{2}\langle\mathcal{R}\rangle_{\mathcal{D}} - \frac{1}{2}\mathcal{Q}_{\mathcal{D}} \label{eq:aD_dot}\\
    0 = \partial_t\langle\rho\rangle_{\mathcal{D}} + 3H_\mathcal{D}\langle\rho\rangle_{\mathcal{D}}\label{eq:continuity}
\end{eqnarray}
where $\langle\rho\rangle_{\mathcal{D}}$, $\langle\mathcal{R}\rangle_{\mathcal{D}}$ and $H_{\mathcal{D}}$ are the averaged matter density, averaged spatial Ricci scalar and the Hubble parameter ($H_{\mathcal{D}}:= \dot{a_{\mathcal{D}}}/a_{\mathcal{D}}$) of the domain $\mathcal{D}$, respectively. $\mathcal{Q}_{\mathcal{D}}$ is called the kinematical backreaction and is defined as 
\begin{equation}\label{eq:Q_D}
    \mathcal{Q}_{\mathcal{D}}:= \frac{2}{3}(\langle\theta^2\rangle_{\mathcal{D}} - \langle\theta\rangle^2_{\mathcal{D}}) - 2\langle\sigma^2\rangle_{\mathcal{D}},
\end{equation}
where $\theta$ is the local expansion rate and $\sigma^2 := 1/2 \sigma_{ij}\sigma^{ij}$ is the squared rate of shear. The Hubble parameter $H_{\mathcal{D}}$ and $\langle\theta\rangle_{\mathcal{D}}$ are related by the relation $H_{\mathcal{D}} = 1/3\langle\theta\rangle_{\mathcal{D}}$. $\mathcal{Q}_{\mathcal{D}}$ is zero for a FLRW-like domain. The necessary condition of integrability connecting (\autoref{eq:aD_ddot}) and (\autoref{eq:aD_dot}) is given by,
\begin{equation}\label{eq:integrability}
    \frac{1}{a^2_{\mathcal{D}}}\partial_t(a^2_{\mathcal{D}}\langle\mathcal{R}\rangle_{\mathcal{D}}) + \frac{1}{a^6_{\mathcal{D}}}\partial_t(a^6_{\mathcal{D}}\mathcal{Q}_{\mathcal{D}}) = 0.
\end{equation}
(\autoref{eq:integrability}) shows an important feature of the averaged equations as it couples the evolution of the averaged intrinsic curvature $(\langle\mathcal{R}\rangle_{\mathcal{D}})$ to the kinematical backreaction term $(\mathcal{Q}_{\mathcal{D}})$ which symbolizes the inclusion of matter inhomogeneities in the analysis. This relationship between $\langle\mathcal{R}\rangle_{\mathcal{D}}$ and $\mathcal{Q}_{\mathcal{D}}$ together with the term $\mathcal{Q}_{\mathcal{D}}$ embodies the diversion from homogeneity.

We now adopt a specific approach within the Buchert formalism, in which ensembles of disjoint regions represent the global domain \cite{Buchert_2015, Buchert4, Weigand_et_al, Rasanen_2004, wiltshire, Kolb_2006, Koksbang_2019, Koksbang2, Koksbang_PRL, Rasanen_2008, bose, Bose2013, Ali_2017, Pandey_2022, Pandey2023, Koksbang5, Koksbang6, Koksbang7, Koksbang8, Halder_2023, Koksbang2023}. Here, the global domain $\mathcal{D}$ is considered to be partitioned into subregions $\mathcal{F}_l$ that themselves consist of elementary space entities $\mathcal{F}_l^{(\alpha)}$. Therefore, mathematically, we have $\mathcal{D} = \cup_l\mathcal{F}_l$, where $\mathcal{F}_l:=\cup_{\alpha}\mathcal{F}^{(\alpha)}_l$ and $\mathcal{F}_l^{(\alpha)}\cap \mathcal{F}_m^{(\beta)} = \emptyset$ for all $\alpha\neq\beta$ and $l\neq m$. 
The average of a scalar-valued function $f$ on the domain $\mathcal{D}$ is given by
\begin{equation}\label{eq:averaging}
\begin{split}
    \langle f\rangle :&= |\mathcal{D}|^{-1}_g \int_\mathcal{D} fd\mu_g
    = \sum_l |\mathcal{D}|_g^{-1}\sum_{\alpha}\int_{\mathcal{F}_l^{(\alpha)}} f d\mu_g\\
    &= \sum_l\frac{|\mathcal{F}_l|_g}{|\mathcal{D}|_g}\langle f\rangle_{\mathcal{F}_l} = \sum_l \lambda_l\langle f\rangle_{\mathcal{F}_l}
\end{split}
\end{equation}
where 
\begin{equation}\label{eq:lambda}
    \lambda_l := \frac{|\mathcal{F}_l|_g}{|\mathcal{D}|_g}
\end{equation}
is the volume fraction of the subregion $\mathcal{F}_l$ such that $\sum_l\lambda_l = 1$ and $\langle f\rangle_{\mathcal{F}_l}$ is the average of $f$ on subregion $\mathcal{F}_l$. The scalar quantities $\rho$, $\mathcal{R}$ and $\mathcal{H}_{\mathcal{D}}$ are governed by (\autoref{eq:averaging}) but $\mathcal{Q}_{\mathcal{D}}$ due to the presence of $\langle\theta\rangle_{\mathcal{D}}^2$ - term, do not adhere to the above equation and instead follow,
\begin{equation}\label{eq:QDsum}
    \mathcal{Q}_{\mathcal{D}} = \sum_l\lambda_l\mathcal{Q}_l + 3\sum_{l\neq m}\lambda_l\lambda_m(H_l-H_m)^2
\end{equation}
where $\mathcal{Q}_l$ and $H_l$ are defined in subregion $\mathcal{F}_l$ in the same way as $\mathcal{Q}_{\mathcal{D}}$ and $H_{\mathcal{D}}$ are defined in the domain $\mathcal{D}$.

We can also define the scale factor $a_l$ for a subregion $\mathcal{F}_l$. By definition, the different subregions are disjoint; therefore it follows that $|\mathcal{D}|_g = \sum_l|\mathcal{F}_l|_g$ and hence, using (\autoref{eq:scale_factor}), we have 

\begin{equation}\label{eq:aD3_sum}
    a^3_{\mathcal{D}} = \sum_l\lambda_{l_i}a_l^3
\end{equation}

where $\lambda_{l_i} = \frac{|\mathcal{F}_{l_i}|_g}{|\mathcal{D}_i|_g}$ is the initial volume fraction which can also be taken as the volume fraction at present and can be represented as $\lambda_{l_0}$, where the subscript $0$ stands for quantities calculated at the present time. Differentiating this relation twice with respect to the foliation time results in,
\begin{equation}\label{eq:aD_sum}
    \frac{\ddot{a}_\mathcal{D}}{a_\mathcal{D}} = \sum_l\lambda_l\frac{\ddot{a}_l(t)}{a_l(t)}+\sum_{l\neq m}\lambda_l\lambda_m(H_l -H_m)^2 \, .
\end{equation}

\section{A model of multiple subregions}\label{sec:our model}

Several studies have been performed within the Buchert averaging scheme using models having just one type of overdense subdomain and one type of underdense subdomain \cite{Koksbang2, Koksbang_PRL,  bose, Bose2013, Ali_2017, Pandey_2022}. This oversimplifies the actual spacetime scenario with matter inhomogeneities where the matter density may vary from very low to very high across different subdomains. Therefore, a more realistic model would have multiple overdense and underdense subregions with distinct evolution profiles. A similar model has been used in \cite{Halder_2023} to study the future evolution of the currently accelerating Universe with multiple inhomogeneous domains.

Here, using Buchert's backreaction framework, we consider a model of the Universe in which domain $\mathcal{D}$ comprises multiple underdense and overdense subregions, {\it viz.}, there are $i$ number of overdense subregions and $i$ number of underdense subregions. The underdense subregions have densities smaller than those of the overdense subregions. Our underdense subregions are modelled to mimic almost empty FLRW regions with very little matter (dust) present. The overdense subregions are modeled to mimic FLRW models with matter (dust) content. The underdense subregions are taken to have Friedmann-like $1/a^2$ negative curvature, while the overdense subregions have Friedmann-like $1/a^2$ positive curvature. The time evolution of the scale factor of $i^{th}$ overdense subregions, $a_{o_i}$ is given in terms of a development angle $\phi_{o_i}$ of the $i^{th}$ overdense subregion \cite{weinberg},
\begin{eqnarray}
    a_{o_i} &=& \frac{q_{o_{i,0}}}{2q_{o_{i,0}} - 1}(1 - \cos{\phi_{o_i}})\label{eq:ao}\\
    t &=& t_0\frac{(\phi_{o_i} - \sin{\phi_{o_i}})}{(\phi_{o_{i,0}} - \sin{\phi_{o_{i,0}}})} \label{eq:ao_t}
\end{eqnarray}
where $q_{o_{i,0}}$ and $\phi_{o_{i,0}}$ are respectively the deceleration parameter of the $i^{th}$ overdense subregion and the value of $\phi_{o_i}$ at time $t_0$,  which is the present time. $q_{o_{i,0}}$ should be greater than $1/2$ \cite{weinberg}. Here, we have taken $q_{o_{i,0}}$ to have a range of values from $1/2$ to $1$. The time $t$ in (\autoref{eq:ao_t}) \cite{Koksbang_2019, Koksbang2023} is the cosmic time, although for each overdense subregion, this $t$ is parameterized in terms of $(\phi_{o_{i,0}}, \phi_{o_i})$ ($\phi_{o_{i,0}}$ itself is a function of $q_{o_{i,0}}$). The value of $t_0$ is the same across all overdense subregions as well as for the global domain which is ensured by the specific form of (\autoref{eq:ao_t}). The time evolution of the scale factor of $i^{th}$ underdense subregions, $a_{u_i}$ is given in terms of a development angle $\phi_{u_i}$ of the $i^{th}$ underdense subregion \cite{weinberg},
\begin{eqnarray}
    a_{u_i} &=& \frac{q_{u_{i,0}}}{1 - 2q_{u_{i,0}}}(\cosh{\phi_{u_i}} - 1)\label{eq:au}\\
    t &=& t_0\frac{(\sinh{\phi_{u_i}} - \phi_{u_i})}{(\sinh{\phi_{u_{i,0}}} - \phi_{u_{i,0}})} \label{eq:au_t}
\end{eqnarray}
where $q_{u_{i,0}}$ and $\phi_{u_{i,0}}$ are respectively the deceleration parameter of the $i^{th}$ underdense subregion and the value of $\phi_{u_i}$ at time $t_0$,  which is the present time. $q_{u_{i,0}}$ has a range of values from $0$ to $1/2$ \cite{weinberg}. The time $t$ in (\autoref{eq:au_t}) is the cosmic time, although for each underdense subregion, this $t$ is parameterized 
in terms of $(\phi_{u_{i,0}}, \phi_{u_i})$ ($\phi_{u_{i,0}}$ itself is a function of $q_{u_{i,0}}$). The value of $t_0$ is the same across all underdense and overdense subregions as well as for the global domain, which is ensured by the specific form of (\autoref{eq:au_t}) and (\autoref{eq:ao_t}). Since the values of $t_0$ and $H_{\mathcal{D}_0}$ are interrelated, one needs to fix either of them \cite{Koksbang2023}. In our subsequent analysis, we choose $H_{\mathcal{D}_0}$ to be 70 $km$ $s^{-1} Mpc^{-1}$. The value of $t_0$ is calculated using the procedure used in \cite{Koksbang2023}, modified for our model (see Appendix \ref{app:t0_calc}).

Note that $a_{o_i}$ and $a_{u_i}$ can be expressed in terms of the volume of the respective subregions using (\autoref{eq:scale_factor}), which gives us
\begin{equation}\label{eq:scale_fact_subreg}
    a_{o_i}(t) := \left(\frac{|\mathcal{D}|_{o_i}}{|\mathcal{D}_0|_{o_i}}\right)^{1/3}; \hspace{0.5 cm} a_{u_i}(t) := \left(\frac{|\mathcal{D}|_{u_i}}{|\mathcal{D}_0|_{u_i}}\right)^{1/3}
\end{equation}
where, $|\mathcal{D}|_{o_i}$ is the volume of the $i^{th}$ overdense subregions at time $t$, $|\mathcal{D}_0|_{o_i}$ is the volume of the $i^{th}$ overdense subregion at time $t_0$ as was done in (\autoref{eq:scale_factor}), and similarly for the case of the underdense subregions.
(\autoref{eq:scale_factor}) and (\autoref{eq:scale_fact_subreg}) require that at $t = t_0$, $a_\mathcal{D}$ = $a_{o_i}$ = $a_{u_i} = 1$, leading to,
\begin{equation}\label{eq:c_values}
    \cos{\phi_{o_{i,0}}} = \left(\frac{1}{q_{o_{i,0}}} - 1\right); \hspace{0.5 cm} \cosh{\phi_{u_{i,0}}} = \left(\frac{1}{q_{u_{i,0}}} - 1\right)
\end{equation}

For a given value of $q_{o_{i,0}}$ and $q_{o_{i,0}}$; $a_{o_i}(t)$ and $a_{u_i}(t)$ can be calculated using (\autoref{eq:ao}), (\autoref{eq:ao_t}), (\autoref{eq:au}), (\autoref{eq:au_t}) and (\autoref{eq:c_values}). Then using (\autoref{eq:aD3_sum}), $a_\mathcal{D}(t)$ can be obtained provided $\lambda_{l_0}$ which is the set of all $\lambda_{{u_i},0}$ and $\lambda_{{o_i},0}$, is known.

It may be noted that $a_\mathcal{D}$ can also be obtained from solving the second order differential equation (\autoref{eq:aD_sum}). Using (\autoref{eq:ao}) and (\autoref{eq:au}) in (\autoref{eq:aD_sum}), we get,
\begin{equation}\label{eq:aD_model}
\begin{split}
\frac{\ddot{a}_{\mathcal{D}}}{a_{\mathcal{D}}}=\left(\sum_i{\lambda_{o_i}\dfrac{\ddot{a}_{o_i}}{a_{o_i}}}\right)+\left(
\sum_j{\lambda_{u_j}\dfrac{\ddot{a}_{u_i}}{a_{u_i}}}\right)\\
+\left(\sum_{k}\sum_{l}{ \lambda_k \lambda_l \left(H_l-H_k\right)^2}\right).
\end{split}
\end{equation}
where, $\lambda_{o_i}$ is the volume fraction of the $i^{th}$ overdense subregion, $\lambda_{u_j}$ is the volume fraction of the $j^{th}$ underdense subregion, $\lambda$ is the set of all $\lambda_{o_i}$ and $\lambda_{u_i}$ and $H$ is respectively, the set of all $H_{o_i}$ and $H_{u_i}$. The combined volume fraction of all the underdense subregions is given by $\lambda_u$, i.e., $\sum_i\lambda_{u_i} = \lambda_u$. Similarly, the total volume fraction of all the overdense subregions is given by $\sum_i\lambda_{o_i} = \lambda_o$. Clearly, $\lambda_o +\lambda_u = 1$. The evaluation of $a_\mathcal{D}$ obtained from these two methods is identical, as confirmed through our analysis.

The volume fraction of the $i^{th}$ overdense subregion can be written as,
\begin{equation}\label{eq:lambda_o_relation}
\begin{split}
    \lambda_{o_i} = \frac{|\mathcal{F}_{o_i}|_g}{|\mathcal{D}|_g}= \frac{a^3_{o_i}|\mathcal{F}_{o_i,0}|_g}{a^3_\mathcal{D}|\mathcal{D}_0|_g} = \lambda_{{o_i},0}\frac{a^3_{o_i}}{a^3_\mathcal{D}}
\end{split}
\end{equation} 
where $t_0$ is a reference time which can be taken as the present time, $|\mathcal{F}_{o_i}|_g$ is the volume of the $i^{th}$ overdense subregion, $|\mathcal{F}_{o_i,0}|_g$ is the volume of the $i^{th}$ overdense subregion at time $t_0$, $|\mathcal{D}_0|_g$ is the volume of the domain $\mathcal{D}$ at time $t_0$ and $\lambda_{{o_i},0}$ is the volume fraction of the $i^{th}$ overdense subregion at time $t_0$. The present time ($t_0$) value of $(\lambda_o,\lambda_u)$ is given by $(\lambda_{o,0},\lambda_{u,0})$ which we have taken to be (0.09,0.91) \cite{Weigand_et_al}.

In our model, we consider the present time volume fraction of $i^{th}$ underdense subregion, $\lambda_{{u_i},0}$ to have a Gaussian distribution within the allowed range of $q_{u_{i,0}}$ from $0$ to $1/2$, given by, 
\begin{equation}
    \lambda_{{u_i},0} = \dfrac{N_u}{\sigma_u \sqrt{2 \pi}} e^{-(q_{u_{i,0}}-\mu_u)^2/2\sigma_u^2},\label{eq:gauss_u}
\end{equation}
where $N_u$ is a normalization constant which ensures that $\sum_i\lambda_{{u_i},0} = \lambda_{u,0} = 0.91$, $\mu_u$ is the mean value of $q_{u_{i,0}}$ and $\sigma_u$ is the standard deviation of $q_{u_{i,0}}$. Therefore, each $i^{th}$ underdense subregion is associated with a particular value of $q_{u_{i,0}}$ and $\lambda_{{u_i},0}$ such that $q_{u_{i,0}}$ varies from $0$ to $1/2$ in the $i$ number of underdense subregions and $\sum_i\lambda_{{u_i},0}  = 0.91$. 

The present-time volume fraction of $i^{th}$ overdense subregion, $\lambda_{{o_i},0}$ is considered to have a Gaussian profile within the allowed range of $q_{o_{i,0}}$ from $1/2$ to $1$ given by, 
\begin{equation}
    \lambda_{{o_i},0} = \dfrac{N_o}{\sigma_o \sqrt{2 \pi}} e^{-(q_{o_{i,0}}-\mu_o)^2/2\sigma_o^2}\label{eq:gauss_o},
\end{equation}
where $N_o$ is a normalization constant which ensures that $\sum_i\lambda_{{o_i},0} = \lambda_{u,0} = 0.09$, $\mu_o$ is the mean value of $q_{o_{i,0}}$ and $\sigma_o$ is the standard deviation of $q_{o_{i,0}}$. In this case, each $i^{th}$ overdense subregion is associated with a particular value of $q_{o_{i,0}}$ and $\lambda_{{o_i},0}$, where $q_{o_{i,0}}$ lies within the range $1/2$ to $1$ across the $i$ number of overdense subregions and $\sum_i\lambda_{{o_i},0}  = 0.09$. The volume fraction of the $i^{th}$ underdense subregion at a time $t$, $\lambda_{u_i}$ is related to the volume fraction at present time $t_0$ by,
\begin{equation}\label{eq:lambda_u_i}
    \lambda_{u_i} = \lambda_{{u_i},0}\left(\dfrac{1-\sum_i \lambda_{o_i}}{1-\sum_i \lambda_{{o_i},0}}\right),
\end{equation}
We have used the Gaussian distribution to define the present time volume fraction of various subregions. The actual physical distribution can only be known by extensive galactic surveys of the matter distribution in the Universe. Although some such surveys have been performed for the local Universe, for the redshifts of our interest, no such surveys exist. Without such surveys, we assume a normal distribution used in analysis where we do not expect any bias. The Gaussian distribution is well-known and extensively used in diverse physical analyses to model unbiased physical conditions. 
(Further details of our model are provided in Appendix \ref{app:model}).

Using (\autoref{eq:QDsum}), the kinematical backreaction term for the domain $\mathcal{D}$ for our model effectively becomes 
\begin{equation}\label{eq:QDsum2}
    \mathcal{Q}_{\mathcal{D}} = \sum_i \lambda_{o_i}\mathcal{Q}_{o_i} + \sum_j \lambda_{u_j}\mathcal{Q}_{u_j} + 3\sum_{l\neq m}\lambda_l\lambda_m(H_l-H_m)^2   \, , 
\end{equation}
where $\mathcal{Q}_{o_i}$ is the kinematical backreaction term for the $i^{th}$ overdense subregion, $\mathcal{Q}_{u_i}$ is for the $i^{th}$ underdense subregion. The summation in the last term runs over the sets of all  $\lambda_{o_i}$, $\lambda_{u_i}$,  $H_{o_i}$ and $H_{u_i}$. (\autoref{eq:integrability}) couples the kinematical backreaction term to the Ricci scalar, and our subregions are also governed by this coupling. Therefore, by selectively choosing the curvatures of our subregions, we can make the respective kinematical backreaction terms for these subregions equal to zero \cite{Weigand_et_al, Rasanen_2006_accelerated}. Hence, in this case, the global kinematical backreaction is governed by only the interplay of the sub-domain Hubble evolutions and volume fractions (third term of (\autoref{eq:QDsum2})). Note that the above assumptions are made in the context of our present model. On the other hand, if the subdomains are endowed with dynamical curvature, other intricate effects could arise through kinematical backreaction, as may also happen in a more general case where the subregions may not necessarily be FLRW. 

Obtaining the values of $\lambda_{o_i,0}$ and $\lambda_{u_i,0}$ from (\autoref{eq:gauss_o}) and (\autoref{eq:gauss_u})  respectively, and using these in (\autoref{eq:lambda_o_relation}) and (\autoref{eq:lambda_u_i}) gives us $\lambda_{o_i}$ and $\lambda_{u_i}$. Hubble parameters for the subregions can be obtained from (\autoref{eq:ao},\autoref{eq:ao_t}) and (\autoref{eq:au}, \autoref{eq:au_t}). We can then use (\autoref{eq:aD_model}) to get $a_\mathcal{D}(t)$ and $H_{\mathcal{D}}(t)$. We next relate these quantities  calculated theoretically from our model with observational quantities (redshift and angular diameter distance by using the covariant scheme \cite{rasanen1, rasanen2}, given by,
\begin{align}
    1+z &= \frac{1}{a_\mathcal{D}}\label{eq:covariant_sch_1}\\
    H_\mathcal{D}\frac{d}{dz}\left((1+z)^2H_\mathcal{D}\frac{dD_A}{dz}\right) &= -4\pi G\langle\rho\rangle_\mathcal{D} D_A.\label{eq:covariant_sch_2}
\end{align}
(\autoref{eq:covariant_sch_1}) relates $a_\mathcal{D}(t)$ with the cosmological redshift $z(t)$ and (\autoref{eq:covariant_sch_2}) relates the angular diameter distance $D_A$ with $\langle\rho\rangle_\mathcal{D}$ and $H_{\mathcal{D}}$. Here, we use (\autoref{eq:covariant_sch_1}) to obtain $z(t)$ from $a_\mathcal{D}(t)$. We can thus evaluate $H_{\mathcal{D}}(z)$ using  $H_{\mathcal{D}}(t)$ (from (\autoref{eq:aD_model})) and $z(t)$ (from (\autoref{eq:covariant_sch_1})).

\begin{figure*}
    \centering
    \begin{tabular}{cc}
	\includegraphics[width=0.5\textwidth]{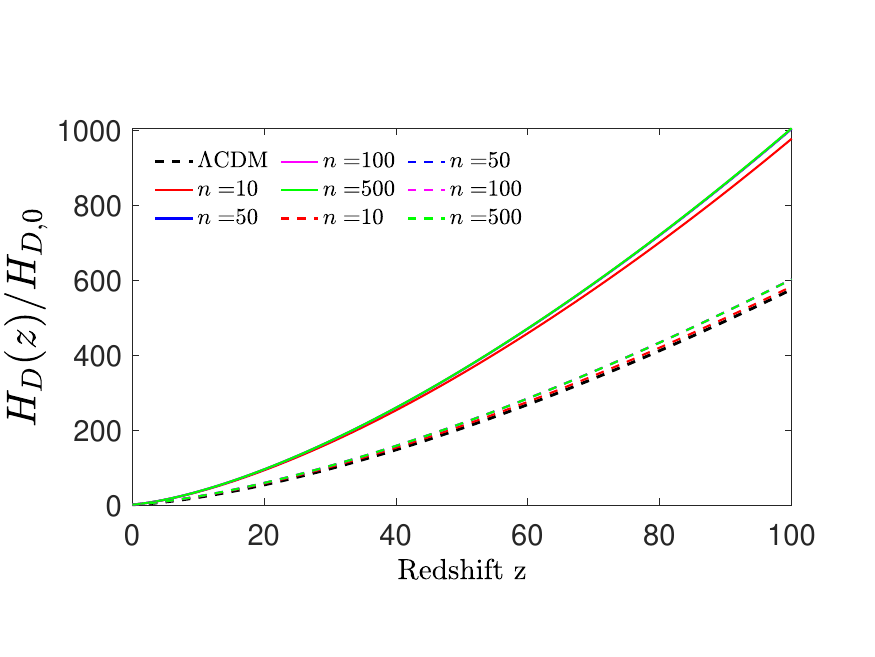}&
	\includegraphics[width=0.5\textwidth]{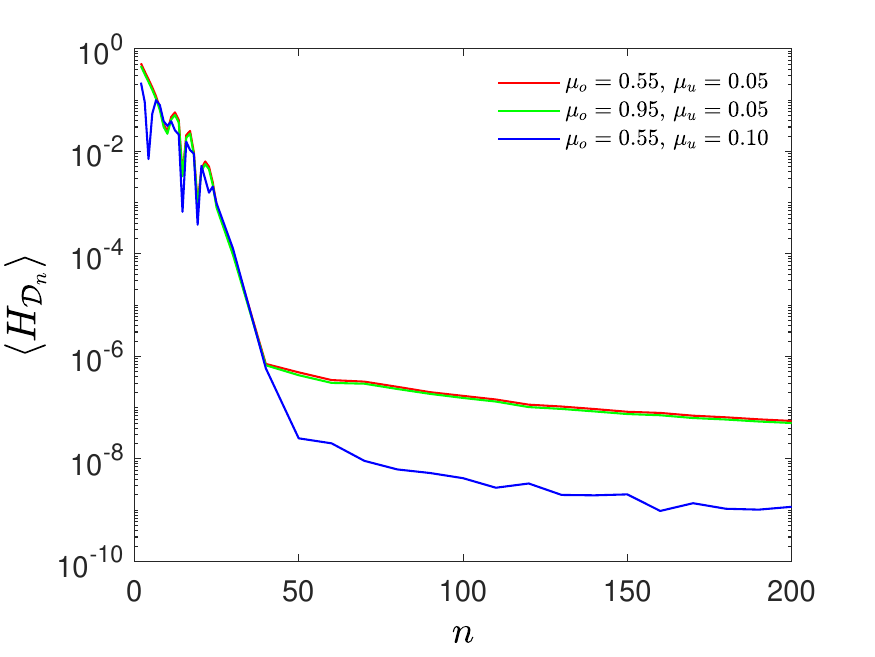}\\
	(a)&(b)\\
    \end{tabular}
    \caption{\label{fig:n_HD} Subplot (a) is the plot of $H_\mathcal{D}(z)/H_{\mathcal{D},0}$ for our backreaction model for different values of $n$, the number of underdense and overdense subregions. Here, $H_{\mathcal{D},0}$ is the value of $H_\mathcal{D}(z)$ at $z = 0$. The values of our model parameters are chosen as follows: $\mu_u = 0.49$, $\sigma_u = 0.01$, $\mu_o = 0.51$ and $\sigma_o = 0.01$ for solid lines and $\mu_u = 0.15$, $\sigma_u = 0.01$, $\mu_o = 0.51$ and $\sigma_o = 0.01$ for dashed lines. The $\Lambda$CDM model curve is shown with a black dashed line. Subplot (b) is the plot of $\left<H_{\mathcal{D}_n} \right>$ versus $n$, the total number of subregions of each type. We take $\sigma_u$ = $\sigma_o$ = 0.01 for all three lines, while $\mu_u$ and $\mu_o$ are varied as mentioned in the legend.}
\end{figure*}

In (\autoref{fig:n_HD}) (a), $H_\mathcal{D}(z)/H_{\mathcal{D},0}$ has been plotted as a function of redshift $z$ for different values of $n$, the number of each type of subregion - overdense and underdense. The total subregions are $2n$, $n$, overdense subregions, and $n$ underdense subregions. Here, $H_{\mathcal{D},0}$ is the value of $H_\mathcal{D}(z)$ at $z = 0$.  Depending on the model parameters, our backreaction model may be very close to a perturbed FRW or mimic a single FRW at very early times. Considering a very high value of $n$ does not lead to a significant difference, as can be seen in our following analysis.
We define
\begin{widetext}
    \begin{equation}
    \left<H_{\mathcal{\mathcal{D}}_n} \right> =  \dfrac{1}{\sum{i}} \sum_i{\left|\frac{\left.\left(H_\mathcal{D}(z)/H_{\mathcal{D},0}\right)\right|_{500}-\left.\left(H_\mathcal{D}(z)/H_{\mathcal{D},0}\right) \right|_{n}}{\left.\left(H_\mathcal{D}(z)/H_{\mathcal{D},0}\right)\right|_{500}} \right|_{z=z_i} }, \label{eq:Hd_avg}\\
\end{equation}
\end{widetext}
which denotes the redshift-averaged variation of $H_{\mathcal{D}_n}$, from the limiting case of $n=500$. In this analysis, we split the redshift range (\emph{i.e.} $0 \leq z\leq 100$) into 100 bins. In (\autoref{eq:Hd_avg}), $i$ is the index number of such bins. The variation of $\left<H_{\mathcal{D}_n}\right>$ with $n$ are plotted in (\autoref{fig:n_HD}(b)). In this figure, the plots are for different chosen sets of $\mu_o$ and $\mu_u$, while the other two parameters are kept at $\sigma_u=0.01$ and $\sigma_o=0.01$. For $n\geq 100$, the average fluctuation is less than $\sim 10^{-6}$. Given the above results, we chose $n=100$ for our remaining calculations.

From here onwards, in our calculations, we consider one hundred under-dense and one hundred over-dense sub-domains. These sub-domains are characterized by the respective volume fractions, $\lambda_{o_i}$ and $\lambda_{u_i}$ (\autoref{eq:lambda_o_relation} and \autoref{eq:lambda_u_i}), distributed using a Gaussian profile among these sub-domains (\autoref{eq:gauss_u} and \autoref{eq:gauss_o}). Our underdense regions are characterized by parameters $q_{u_{i,0}}$ that vary from $0 < q_{u_{i,0}} < 0.5$ \cite{weinberg}. This range for $q_{u_{i,0}}$ has been taken to ensure a wide range of underdense subregions is present in our model to mimic a variety of underdense regions that may be present in the Universe. $\mu_u$ and $\sigma_u$ are the mean and standard deviation of the Gaussian profile of the underdense regions. The underdense subregion with $q_{u_{i,0}} = \mu_u$ will have the largest value of $\lambda_{u_i}$ and thus will be the most prominent underdense subregion in the analysis. Here, $\sigma_u$ governs the distribution width about a given $\mu_u$. Similarly, our overdense regions are characterized by parameters $q_{o_{i,0}}$ varying from $1/2 < q_{o_{i,0}} < 1$. This range for $q_{o_{i,0}}$ has been taken to ensure that a wide range of overdense subregions is present in our model to mimic a variety of overdense regions that may be present in the Universe. $\mu_o$ and $\sigma_o$ are the mean and standard deviation for the Gaussian profile of overdense regions. The overdense subregion with $q_{o_{i,0}} = \mu_o$ will have the highest value of $\lambda_{o_i}$ and, therefore, will be the most prominent overdense subregion in the analysis. $\sigma_o$ is the standard deviation of the distribution, which governs the width of the distribution about the mean value. 

\begin{figure}
    \centering
    \includegraphics[width = 0.5\textwidth]{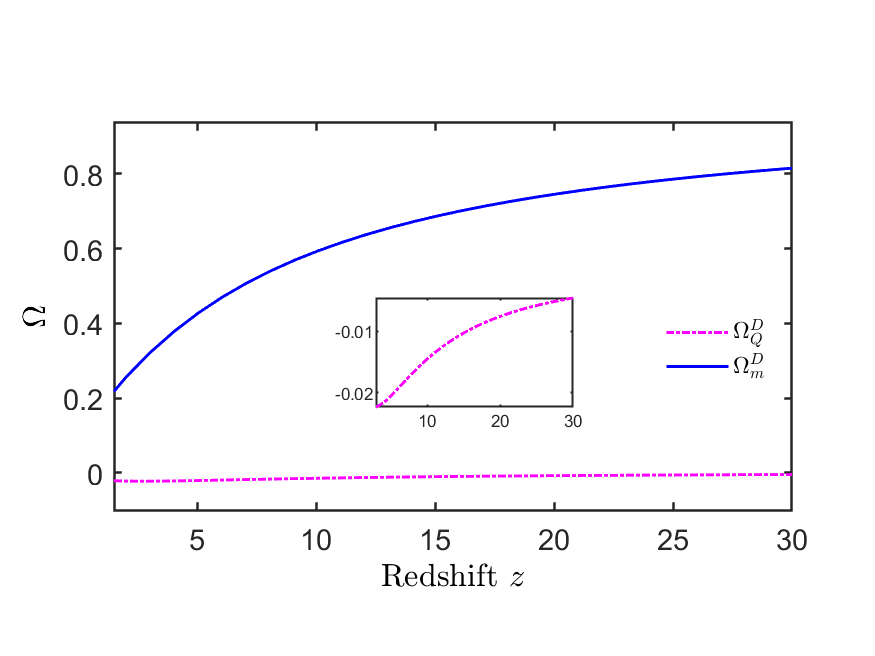}
    \caption{Plot of average density parameters, $\Omega^\mathcal{D}_m$ and $\Omega^\mathcal{D}_\mathcal{Q}$ as a function of redshift, $z$. The values of the parameters are chosen as: $\mu_u = 0.01$, $\sigma_u = 0.01$, $\mu_o = 0.99$ and $\sigma_o = 0.01$. The inset shows the magnified plot for $\Omega^{\mathcal{D}}_\mathcal{Q}$, the density parameter for kinematical backreaction term $\mathcal{Q}_{\mathcal{D}}$.}\label{fig:omega}
\end{figure}

In (\autoref{fig:omega}), the average density parameters are plotted as a function of the redshift in the redshift range $(z<30)$. (See Appendix (\ref{app:model}) for the calculation of these average density parameters). The density parameter associated with kinematical backreaction $\mathcal{Q}_\mathcal{D}$ plays a significant role in our backreaction model at late redshifts (around $z=5$, as seen from the inset), embodying the departure from FLRW behavior in our framework. The $\mathcal{Q}_\mathcal{D}$ term becomes negligible at large redshifts,  which can be seen from the corresponding plot in (\autoref{fig:omega}). The term $\Omega^{\mathcal{D}}_\mathcal{Q}$ going to zero at high redshifts (or at early times) shows that our model can mimic a perturbed FLRW model at early times.

\begin{figure*}
    \centering
    \begin{tabular}{cc}
	\includegraphics[width=0.5\textwidth]{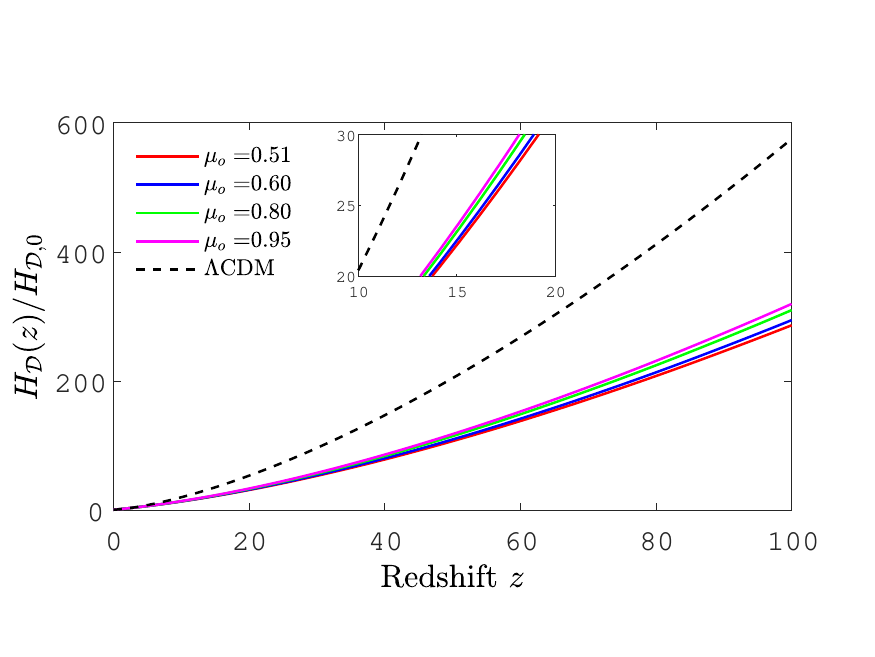}&
	\includegraphics[width=0.5\textwidth]{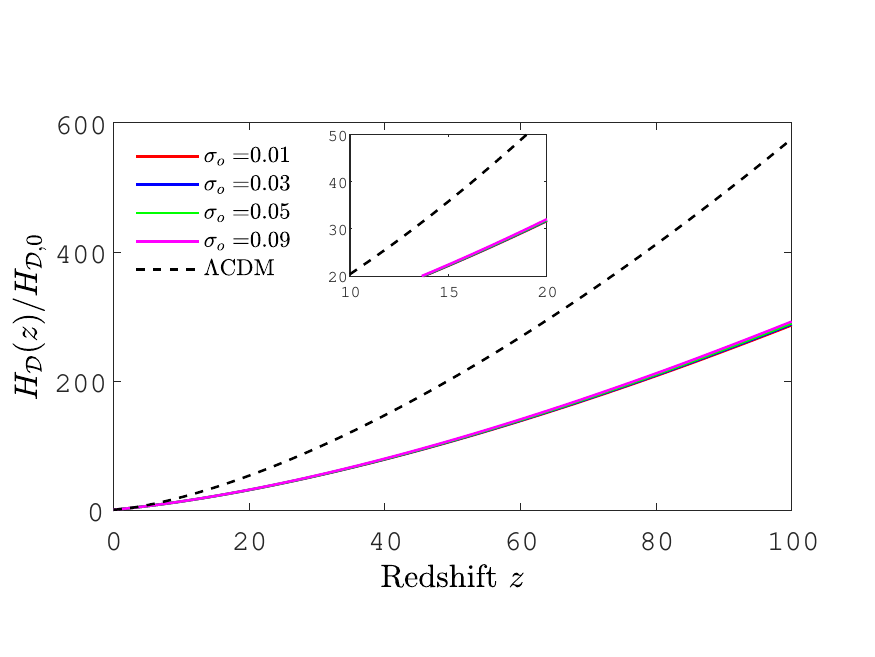}\\
	(a)&(b)\\
	\includegraphics[width=0.5\textwidth]{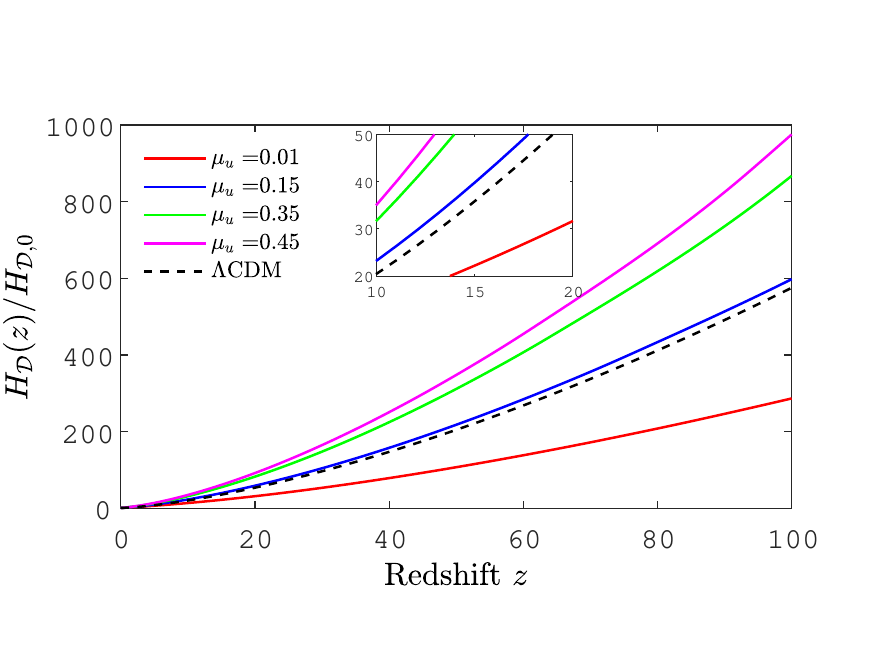}&
	\includegraphics[width=0.5\textwidth]{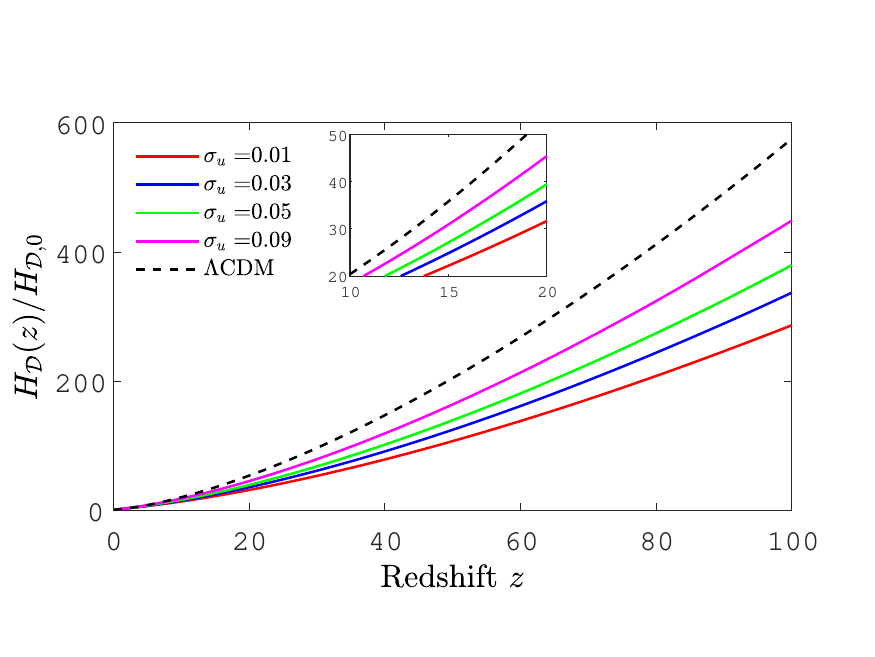}\\
	(c)&(d)\\
    \end{tabular}
    \caption{\label{fig:H} Plots of $H_\mathcal{D}(z)/H_{\mathcal{D},0}$ for $\Lambda$CDM and our backreaction model as a function of z. Our backreaction model has four parameters that can be varied: $\mu_u, \sigma_u, \mu_o, \sigma_o$. In (a) $\mu_o$ is varied with $\sigma_o = \sigma_u = 0.01$ and $\mu_u = 0.01$ fixed. In (b) $\sigma_o$ is  varied with $\mu_o = 0.51, \sigma_u = 0.01$ and $\mu_u = 0.01$ fixed. In (c), $\mu_u$ is varied with $\sigma_o = \sigma_u = 0.01$ and $\mu_o = 0.51$ fixed. In (d), $\sigma_u$ is varied with $\sigma_o = 0.01, \mu_o = 0.51$ and $\mu_u = 0.01$ fixed. The insets show the plot lines for the redshift range 10-20. The value of $H_0$ (Hubble parameter at the present time) used is 100 $h$ km $\rm{s^{-1} Mpc^{-1}}$ where $h = 0.7$.}
\end{figure*}

In (\autoref{fig:H}), the variation with respect to the redshift of $H_\mathcal{D}(z)/H_{\mathcal{D},0}$ (here, $H_{\mathcal{D},0}$ is the value of $H_\mathcal{D}(z)$ at $z = 0$) for our backreaction model and for the $\Lambda$CDM model ($H(z)/H_0$ for $\Lambda$CDM, $H_0 = H(z=0)$) are plotted.  Our backreaction model has four parameters that can be varied: $\mu_u$, $\sigma_u$, $\mu_o$, $\sigma_o$. One of the above parameters is varied in the four sub-figures while keeping the other three fixed. From subplots (a), it can be observed that larger values of parameter $\mu_o$ result in larger values of the quantity $H_\mathcal{D}(z)/H_{\mathcal{D},0}$, although the variation is very less and from (c) also, it can be observed that larger values of parameter $\mu_u$ result in larger values of the quantity $H_\mathcal{D}(z)/H_{\mathcal{D},0}$. Let us first consider the subplot (c) where $\mu_u$ is being varied, keeping the other three parameters fixed. Since $\mu_u$ is the mean of the Gaussian distribution of $\lambda_{{u_i},0}$, it corresponds to the subregion with the largest value of $\lambda_{{u_i},0}$ in the distribution. The subregion with $q_{u_{i,0}}$ = $\mu_u$ possesses the largest value of $\lambda_{{u_i},0}$ and therefore, the largest value of $\lambda_{u_i}$ also, from (\autoref{eq:lambda_u_i}). It follows from (\autoref{eq:averaging}) that this underdense subregion through its $H_{u_i}$ provides the largest contribution among all other underdense subregions in the determination of $H_u$, the total Hubble parameter for all the underdense subregions combined, and consequently, provides the largest contribution in $H_{\mathcal{D}}$ among all other underdense subregions. Therefore, in subplot (c) with all other parameters fixed, the plotlines for $H_{\mathcal{D}}$  follow the trend of variation of $H_{u_i}$ with respect to the redshift $z$, of the subregion with $q_{u_{i,0}}$ = $\mu_u$. The higher values of $q_{u_{i,0}}$ result in higher values of $H_u(z)$ at higher values of $z$, which is observed in subplot (c), where higher values of $\mu_u$ give higher values of $H_\mathcal{D}(z)/H_{\mathcal{D},0}$. The behavior of the plotlines in subplot (a) can also be explained similarly, where the corresponding underdense subregion analysis replaces the overdense subregion analysis. In subplot (a), there is not much difference between the plotlines. This can be ascertained to the fact that overdense subregions have less impact on the global domain dynamics. The reason for this is that the collective volume fraction of the overdense subregions is much smaller than the collective volume fraction of the underdense subregions.  

 \par On the other hand, from the subplots (b) and (d), it can be seen that larger values of $\sigma_o$ and $\sigma_u$ lead to larger values of the quantity $H_\mathcal{D}(z)/H_{\mathcal{D},0}$. Note that $\sigma_o$ and $\sigma_u$ represent the spread of the Gaussian distributions. In subplots (b) and (d), only $\sigma_o$ and $\sigma_u$ are varied respectively, keeping the other three parameters fixed. Therefore, a wider distribution with the same mean is considered in subplots (b) and (d). A wider distribution results in more subregions becoming significant than for a narrower distribution. As the contributions of more subregions become effective, the values of the combined Hubble parameters for the overdense and underdense subregions, $H_o$ and $H_u$, respectively, increase, and hence the value of $H_{\mathcal{D}}$ also increases, which is observed in subplots (b) and (d). Similar to the case of subplot (a), varying $\sigma_o$ in subplot (b) does not have much effect on the plotlines. Therefore, model parameters associated with the overdense subregions do not have significant impact on the global domain dynamics.


\section{\label{sec:analysis} Effect of inhomogeneities on the 21 cm brightness temperature}

To analyze the brightness temperature of the 21-cm signal in the context of our model of multiple subregions of spacetime with matter distribution inhomogeneities, we replace $H(z)$ in the equations of 21-cm cosmology with $H_{\mathcal{D}}(z)$ (\autoref{fig:H}) which is the effective Hubble parameter calculated from our model using (\autoref{eq:aD_model}).
Here, we employ the general scheme to calculate the 21-cm brightness temperature $T_{21}$ for both $\Lambda$CDM and our backreaction model. The only difference between these two models is calculating the Hubble parameter $H(z)$, where $z$ is the redshift. For the $\Lambda$CDM model, $H(z)$ is calculated using the standard relation of the Hubble parameter with various density parameters, $\Omega$s. In contrast, for our model $H(z)$ is replaced by $H_{\mathcal{D}}(z)$, since we are interested in the evaluation of all physical quantities with respect to the global domain.

From (\autoref{eq:T21}), using Taylor expansion of $e^{-\tau(z)}$, and
ignoring higher  order terms of $\tau(z)$, we get 
\begin{equation*}
    T_{21} \approx \frac{T_s - T_\gamma}{1+z}\tau(z)
\end{equation*}
Now, from (\autoref{eq:tau}), for $\delta_rv_r = 0$, $\tau(z) \propto 1/H(z)$. Therefore,
\begin{equation}
    T_{21} \propto \frac{T_s - T_\gamma}{1+z}\frac{1}{H(z)} \label{eq:t21hz}
\end{equation}
Thus, for a given value of $T_s$ and $T_\gamma$, $T_{21}$ is inversely proportional to $H(z)$. Note that the sign of $T_{21}$ is governed by $T_s$ and $T_\gamma$. If $T_s > T_\gamma$, then $T_{21}$ is positive and negative for vice versa. $H(z)$ has effect only on the magnitude of $T_{21}$. Also note that $T_{21}$ is related to $H(z)$ via $T_s$, which itself depends on $H(z)$ (from (\autoref{eq:Ts2}) and (\autoref{eq:Tb})), but the dominant and more direct relationship between $T_{21}$ and $H(z)$ is from (\autoref{eq:t21hz}). 

\begin{figure*}
    \centering
    \begin{tabular}{cc}
	\includegraphics[width=0.5\textwidth]{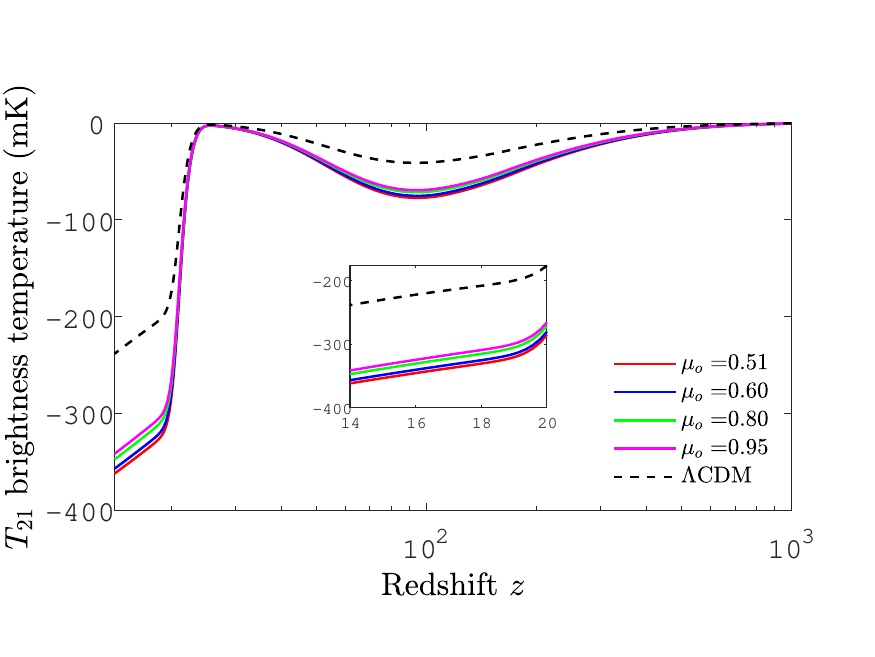}&
	\includegraphics[width=0.5\textwidth]{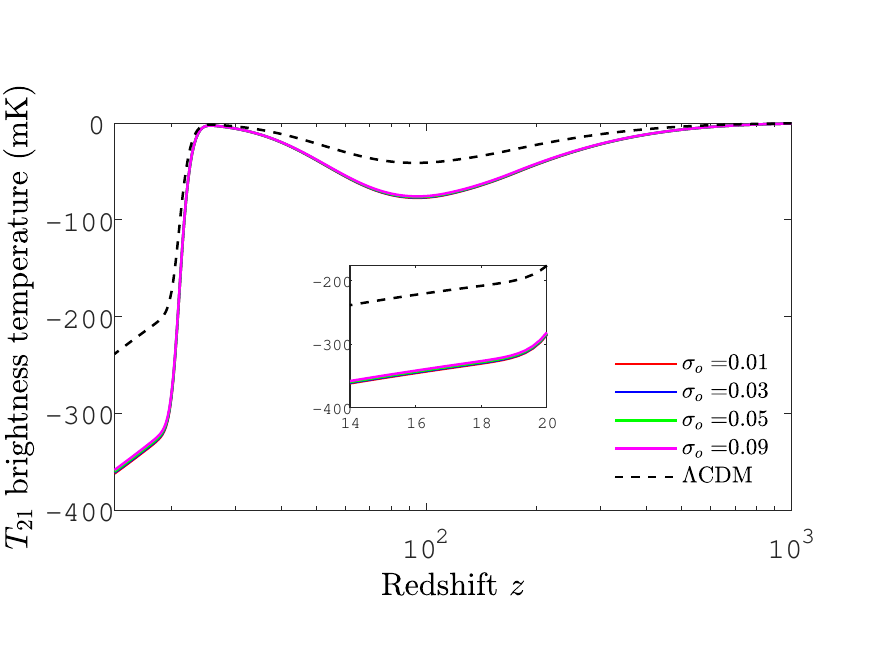}\\
	(a)&(b)\\
	\includegraphics[width=0.5\textwidth]{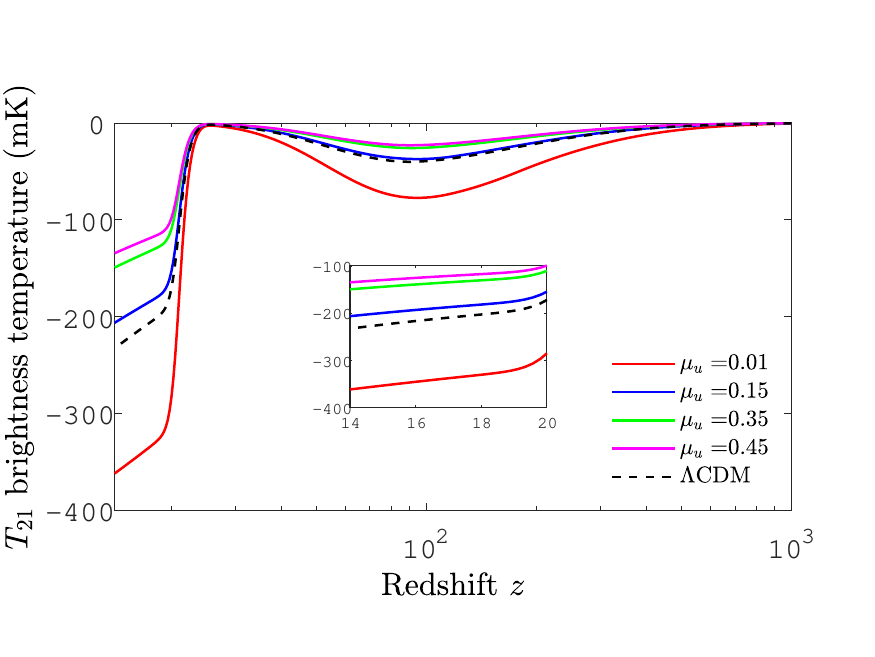}&
	\includegraphics[width=0.5\textwidth]{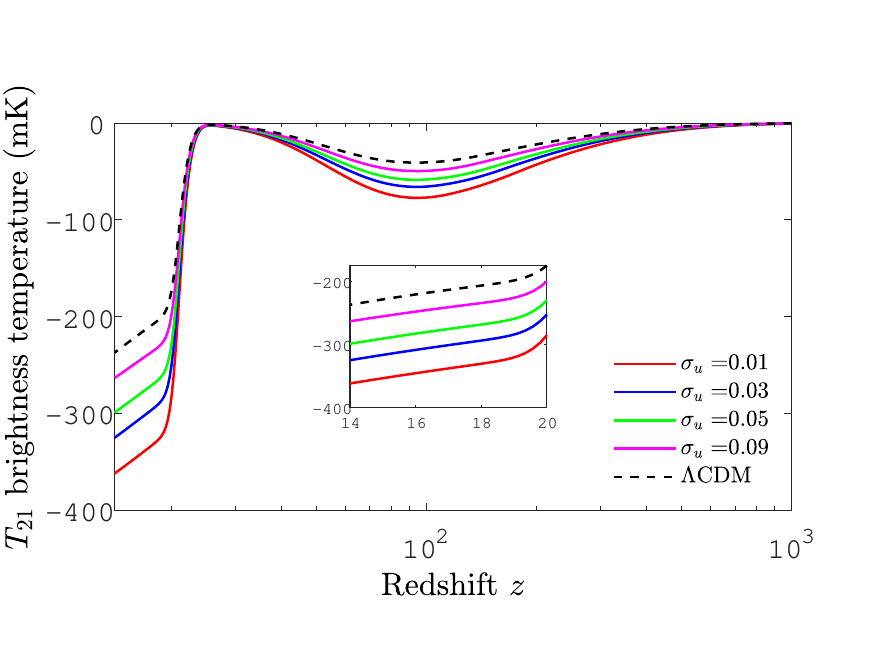}\\
	(c)&(d)\\
    \end{tabular}
    \caption{\label{fig:T21} Plots of brightness temperature $T_{21}$ for the $\Lambda$CDM model and our backreaction model for the redshift range $14 - 1000$. Our backreaction model has four parameters that can be varied: $\mu_u, \sigma_u, \mu_o, \sigma_o$. In (a) $\mu_o$ is varied with $\sigma_o = \sigma_u = 0.01$ and $\mu_u = 0.01$ fixed. In (b) $\sigma_o$ is  varied with $\mu_o = 0.51, \sigma_u = 0.01$ and $\mu_u = 0.01$ fixed. In (c), $\mu_u$ is varied with $\sigma_o = \sigma_u = 0.01$ and $\mu_o = 0.51$ fixed. In (d), $\sigma_u$ is varied with $\sigma_o = 0.01, \mu_o = 0.51$ and $\mu_u = 0.01$ fixed. The value of $H_0$ (Hubble parameter at the present time) used is 100 $h$ km $\rm{s^{-1} Mpc^{-1}}$ where $h = 0.7$. The insets show the plot lines in the redshift range of our interest.}
\end{figure*}

Though the primary redshift range of interest for our present analysis is $14<z<20$ corresponding to the range of the EDGES result \cite{EDGES_Bowman2018}, there are various future proposed experiments to analyze the 21 cm signal at various redshift ranges \cite{Burns_2017, DARE, Pratush, REACH, BIGHORNS}. In \autoref{fig:T21}, we display results for a large redshift range up to $z = 1000$, given the above-proposed observations. The current analysis focuses on a much narrower regime of $14<z<20$ as displayed in the figure insets.

In (\autoref{fig:T21}), the variations of the brightness temperature $T_{21}$ in mK as a function of redshift $z$ (using (\autoref{eq:T21})) in the redshift range $14 - 1000$ are plotted for both the $\Lambda$CDM model and our backreaction model. In the case of our model, $H(z)$ is replaced by $H_{\mathcal{D}}(z)$ (plotted in (\autoref{fig:H})). Our backreaction model has four parameters that can be varied: $\mu_u, \sigma_u, \mu_o, \sigma_o$. One of the parameters varied in each of the four subfigures, while the other three were fixed. Each subplot of (\autoref{fig:T21}) has a relation with the corresponding subplot of (\autoref{fig:H}) via (\autoref{eq:t21hz}). In subplot (c) of (\autoref{fig:H}), lower values of $\mu_u$ yielded lower values of $H_{\mathcal{D}}(z)$, and since $H_{\mathcal{D}}(z)$ is inversely proportional to the magnitude of $T_{21}$, lower values of $\mu_u$ should give us the greater magnitude of $T_{21}$. This is what is observed in subplot (c) of (\autoref{fig:T21}). Other subplots of (\autoref{fig:T21}) also have a one-to-one correspondence with their counterparts in (\autoref{fig:H}), which can be explained similarly.
At lower values of the redshift $z$,  our backreaction model for a large range of parameters leads to lower brightness temperature than the $\Lambda$CDM model. In general, lower values of $\mu_o$, $\mu_u$, $\sigma_o$ and $\sigma_u$ lead to lower (more negative) values of $T_{21}$. 

\begin{figure*}
    \centering
    \begin{tabular}{cc}
	\includegraphics[trim={0 0 85 0},clip, width=0.5\textwidth]{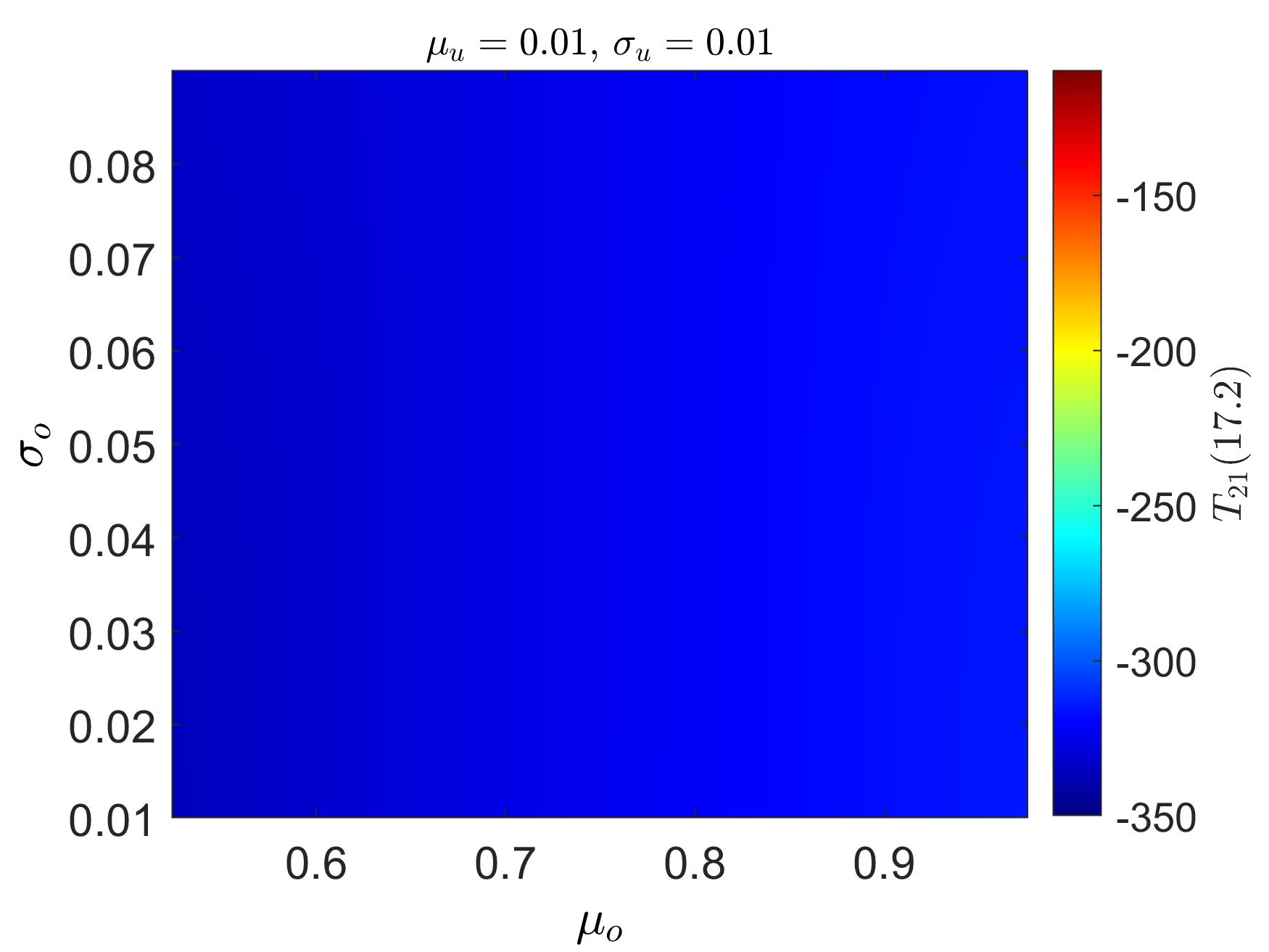}&
	\includegraphics[trim={0 0 85 0},clip, width=0.5\textwidth]{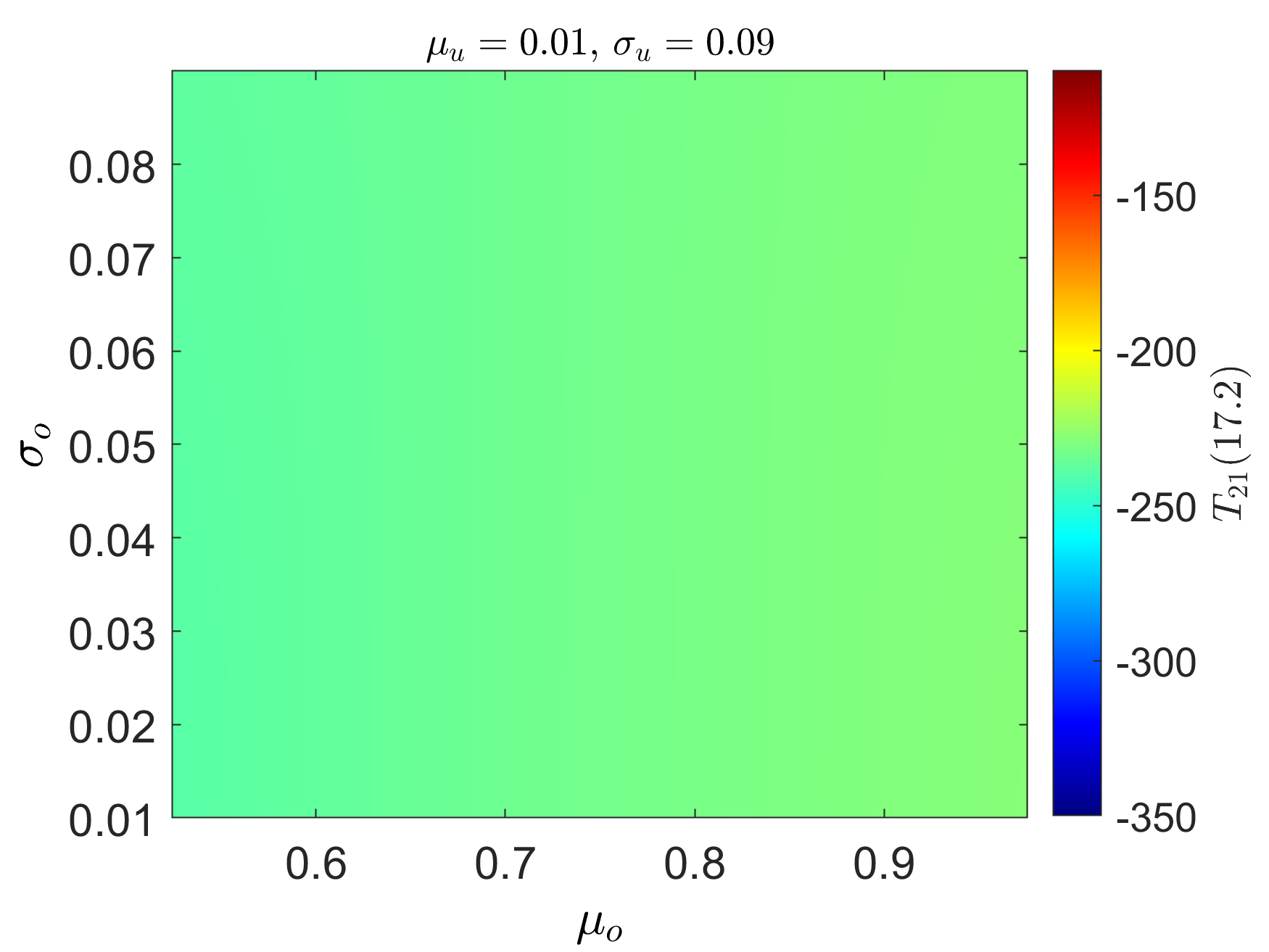}\\
	(a)&(b)\\
	\includegraphics[trim={0 0 85 0},clip, width=0.5\textwidth]{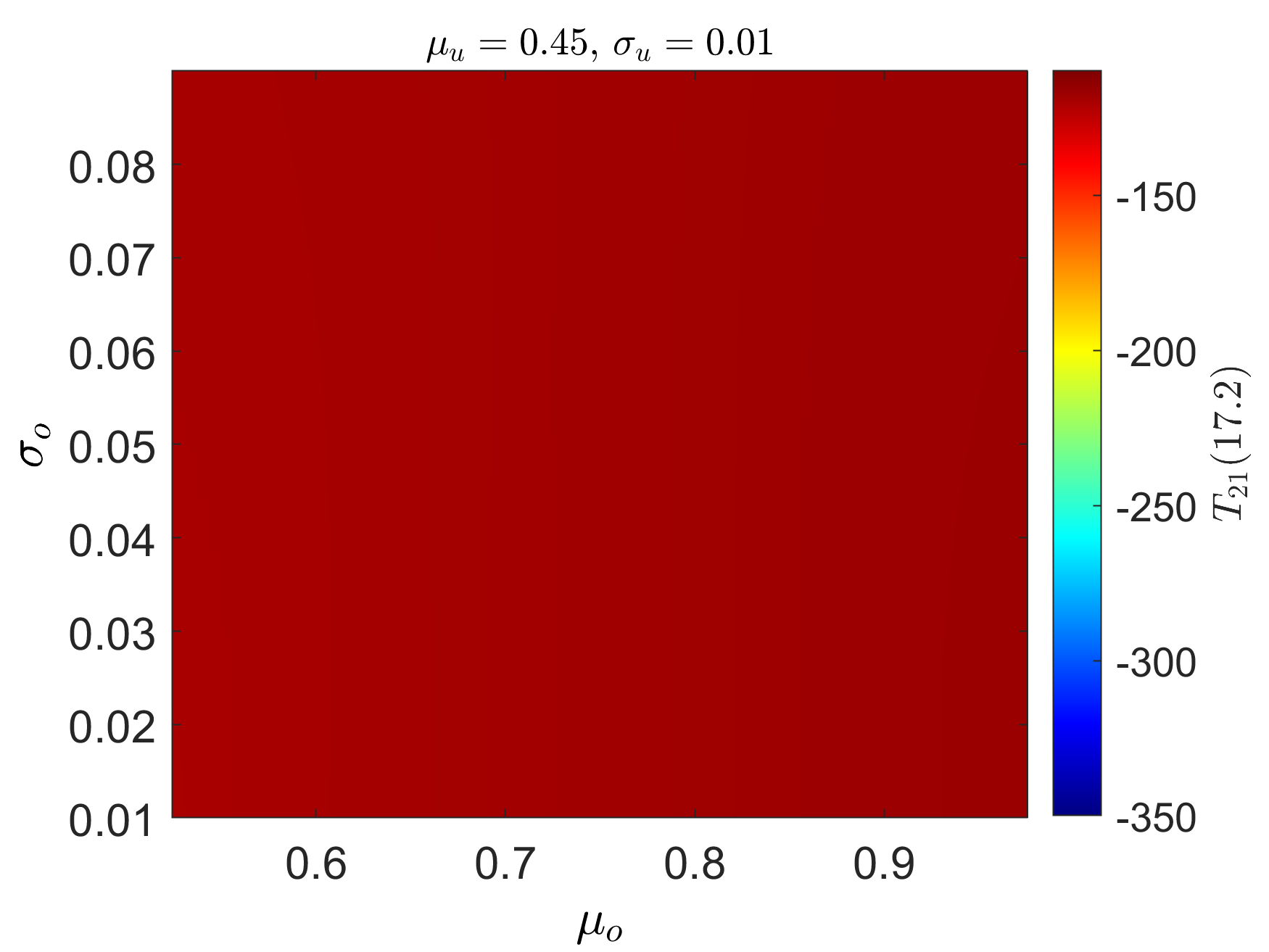}&
	\includegraphics[trim={0 0 85 0},clip, width=0.5\textwidth]{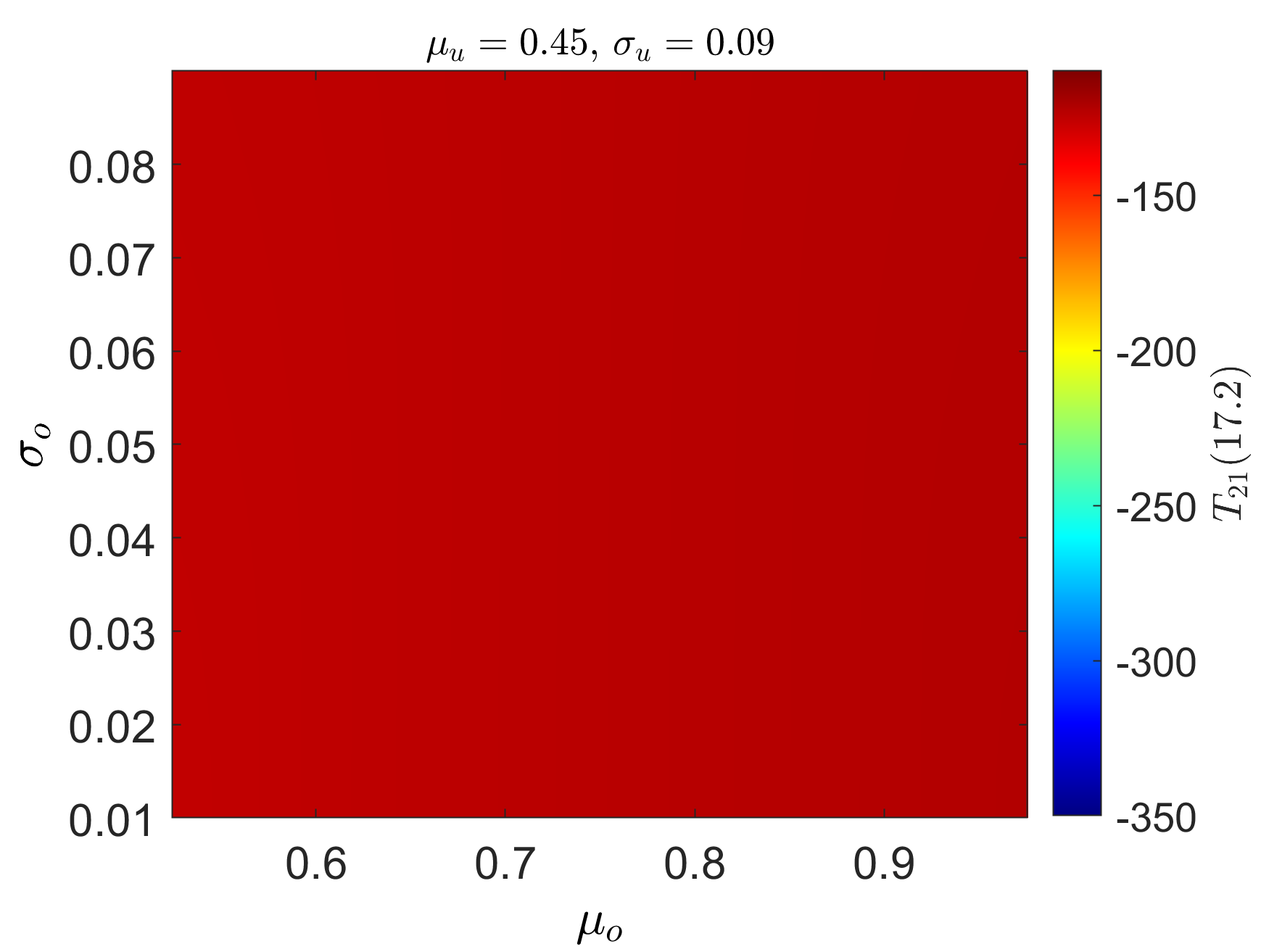}\\
	(c)&(d)\\
    \end{tabular}
    \begin{tabular}{c}
	\includegraphics[trim={0 0 0 230},clip, width=0.7\textwidth]{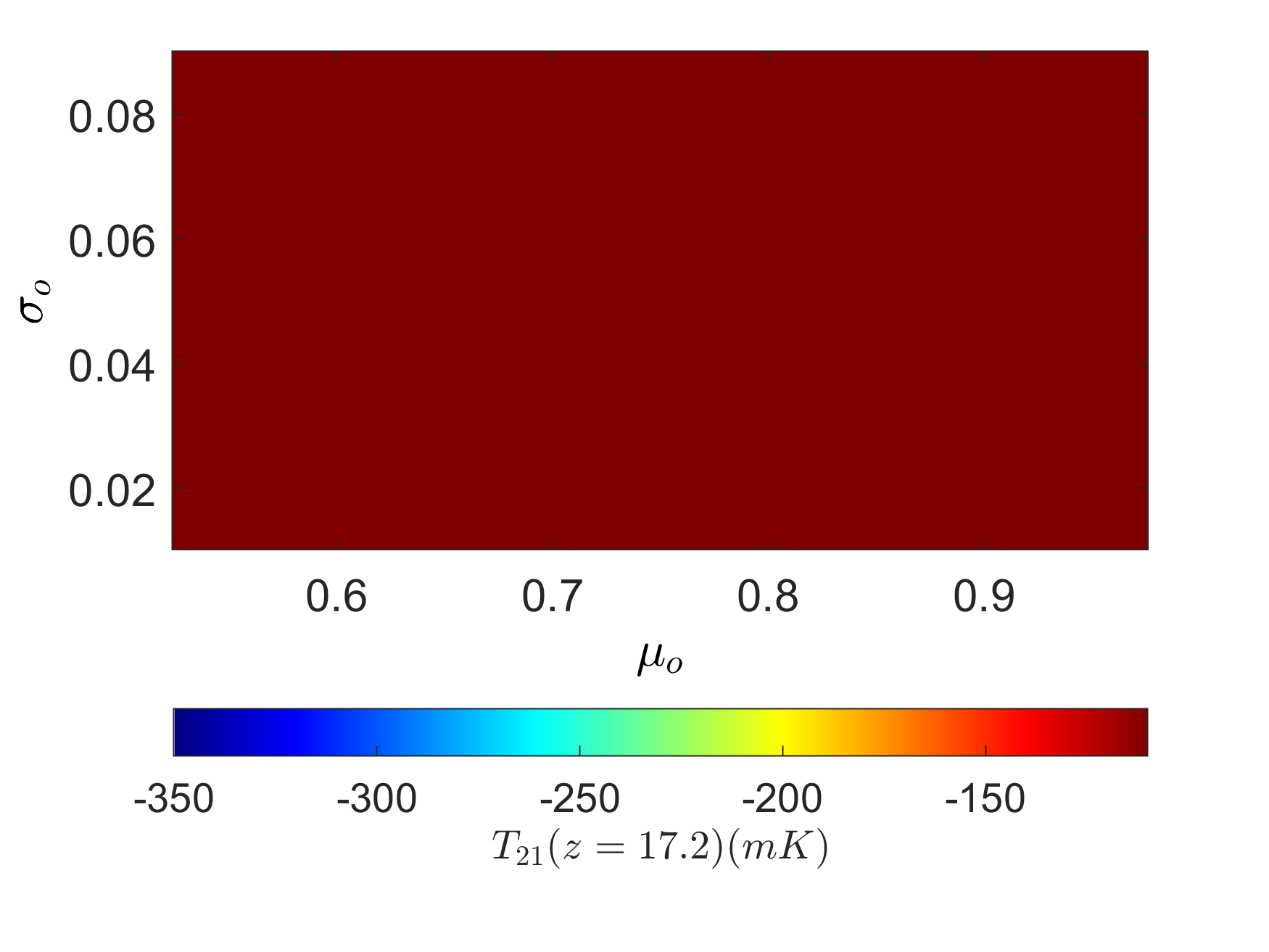}\\		
    \end{tabular}
    \caption{\label{fig:T21_cont_o} Contour representation of $T_{21} (z=17.2) (mK)$ in the $\sigma_o - \mu_o$ plane for (a) $\mu_u=0.01$, $\sigma_u=0.01$ (b) $\mu_u=0.01$, $\sigma_o=0.09$, (c) $\mu_u=0.45$, $\sigma_u=0.01$, (d) $\mu_u=0.45$, $\sigma_u=0.09$.}
\end{figure*}

In (\autoref{fig:T21_cont_o}), the variation of $T_{21}$ at $z = 17.2$ in the $\mu_o-\sigma_o$ plane is shown for different sets of values of $(\mu_u,\sigma_u)$ using a contour plot. The value of $\mu_o$ varies in the range of $0.5 - 1.0$ along the x-axis, while $\sigma_o$ is varied in the range $0.01 - 0.09$ along the y-axis. The contour colors describe the value of $T_{21}(z=17.2)$ per the color bar at the bottom of the figure. In subplots (a) and (b), $(\mu_u,\sigma_u)$ are $(0.01,0.01)$ and $(0.01,0.09)$, respectively.  Subplot (a) has a lower value (more negative) of $T_{21}(z=17.2)$ compared to subplot (b). There is also very little variation within the individual  subplots (a) and (b). This shows that fixing $(\mu_u,\sigma_u)$ and varying $(\mu_o,\sigma_o)$ has very little effect on the calculation of $T_{21}$ at $z = 17.2$. However, changing $\sigma_u$ between subplots (a) and (b) results in significant variation. This also shows the insignificance of the model parameters associated with overdense subregions in the dynamics of the global domain. From subplots (c) and (d), it can be seen that $T_{21}(z=17.2)$ has high values (from -150 mK to -110 mK) for $(\mu_u,\sigma_u) = (0.45,0.01)$ and $(0.45,0.09)$. Changing $\sigma_u$ from 0.01 to 0.09 while keeping $\mu_u$ fixed in (c) and (d) has little effect on the value of $T_{21}(z=17.2)$. In these subplots, $T_{21}(z=17.2)$ has the lowest value of around $-350$ mK in the lower left portion of the subplot (a). This affirms our analysis of (\autoref{fig:T21}) that lower values of $\sigma_u$, $\sigma_o$, $\mu_o$ and $\mu_u$ lead to lower values (more negative) of $T_{21}(z=17.2)$. From subplots (a) and (c), it can be seen that changing the value of $\mu_u$ keeping $\sigma_u$ fixed has a more prominent effect on the brightness temperature than doing vice versa. 
 
\begin{figure*}
    \centering
    \begin{tabular}{cc}
	\includegraphics[trim={0 0 85 0},clip, width=0.5\textwidth]{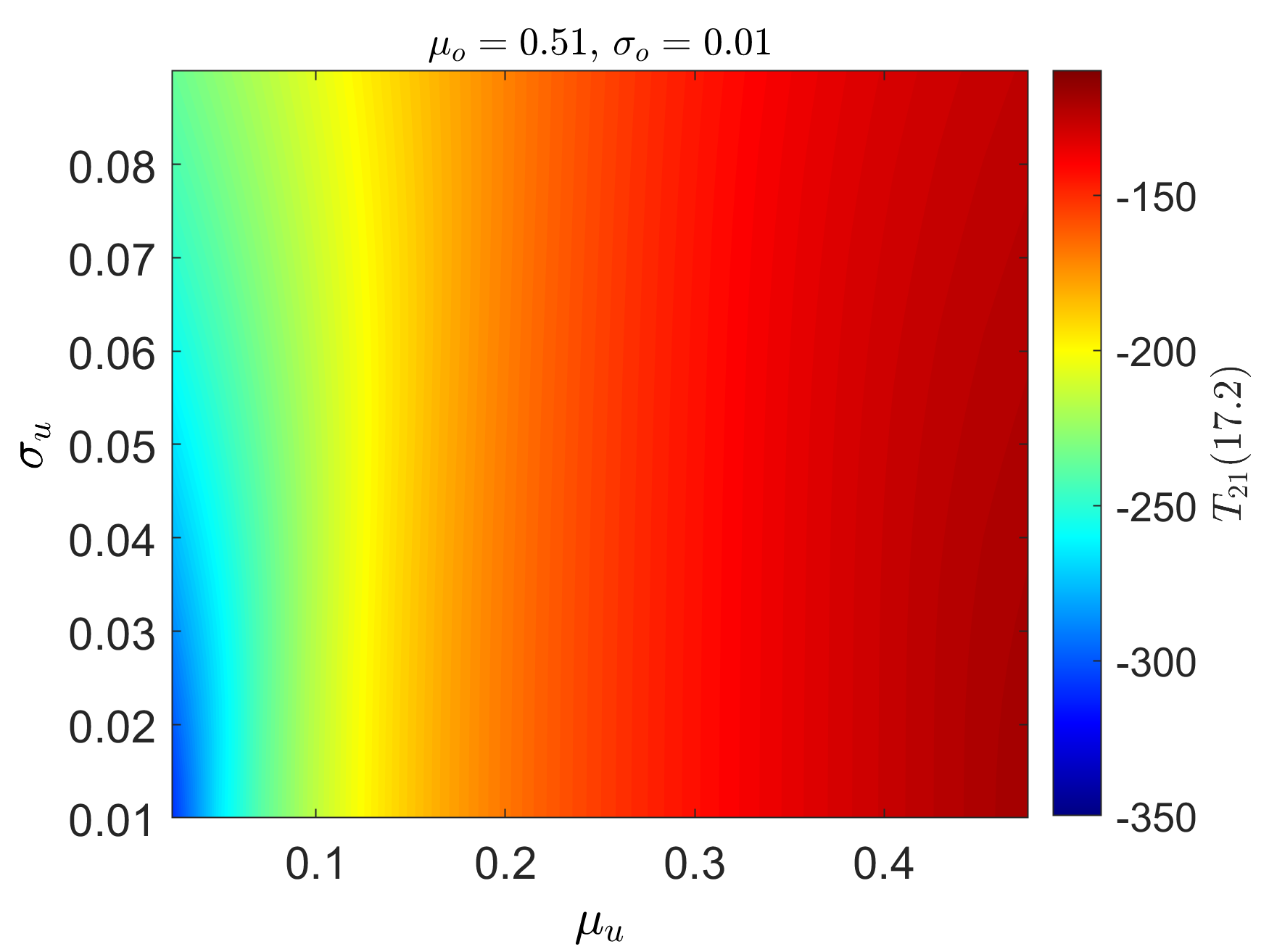}&
	\includegraphics[trim={0 0 85 0},clip, width=0.5\textwidth]{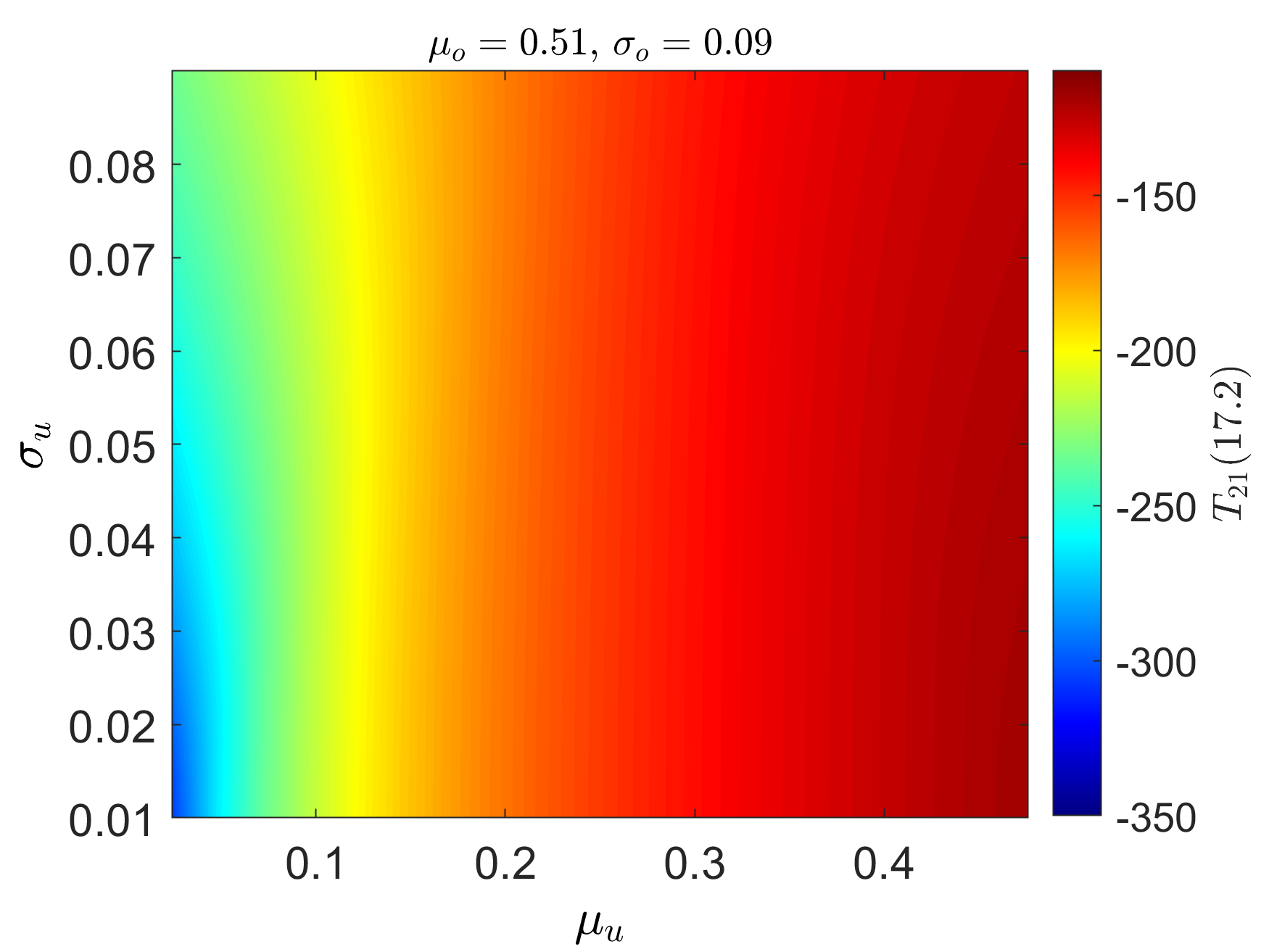}\\
	(a)&(b)\\
	\includegraphics[trim={0 0 85 0},clip, width=0.5\textwidth]{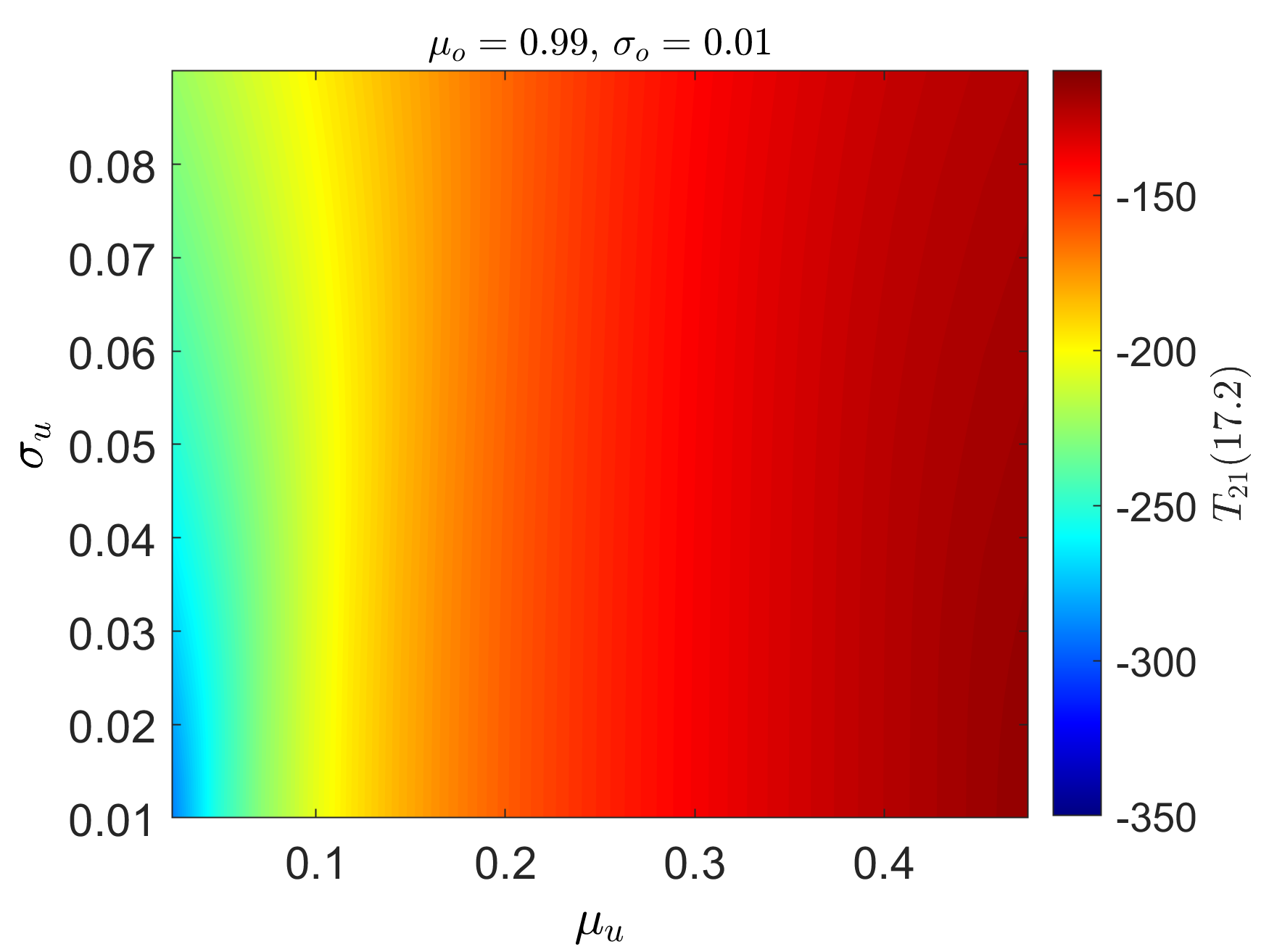}&
	\includegraphics[trim={0 0 85 0},clip, width=0.5\textwidth]{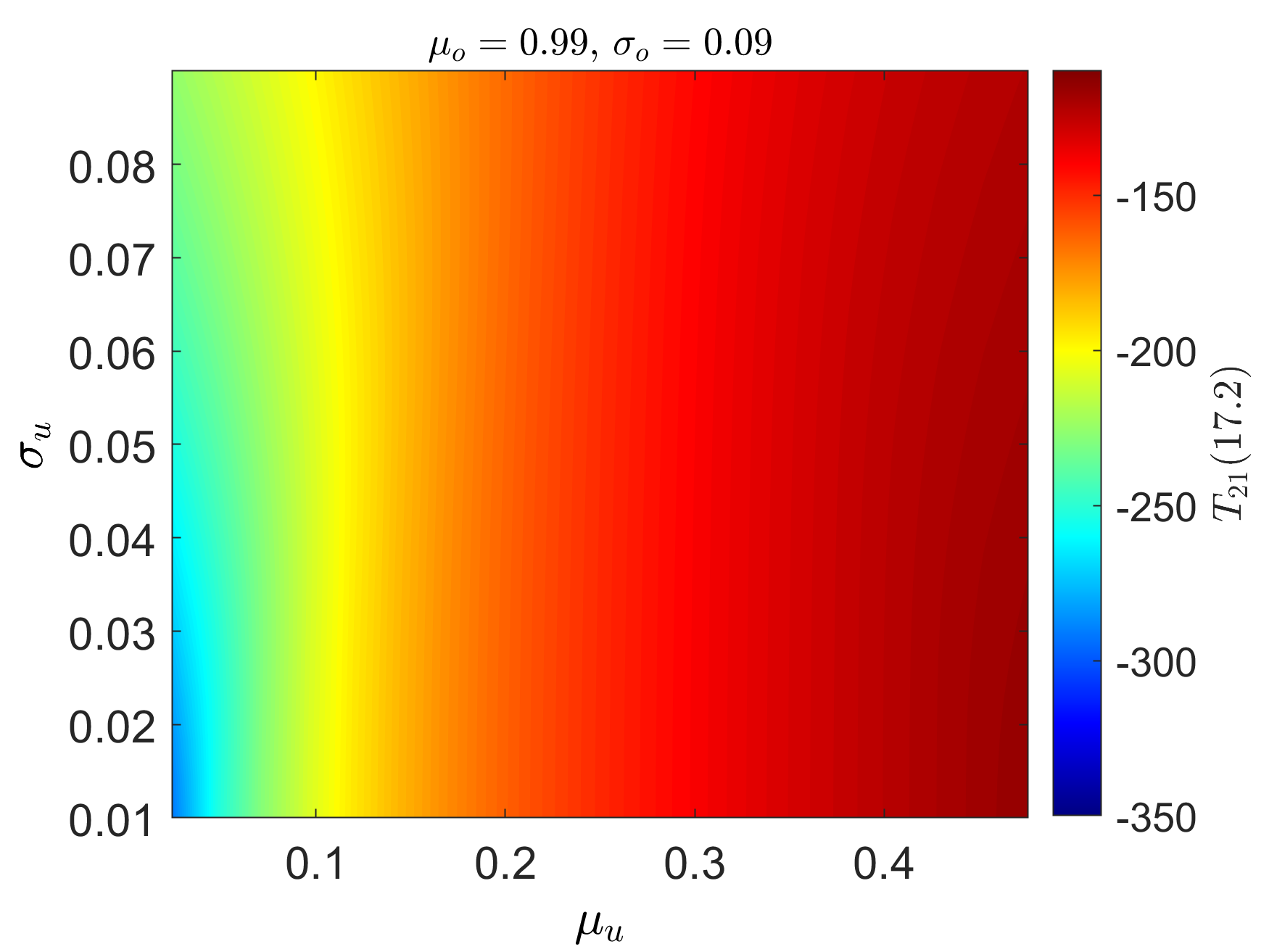}\\
	(c)&(d)\\
    \end{tabular}
    \begin{tabular}{c}
	\includegraphics[trim={0 0 0 230},clip, width=0.7\textwidth]{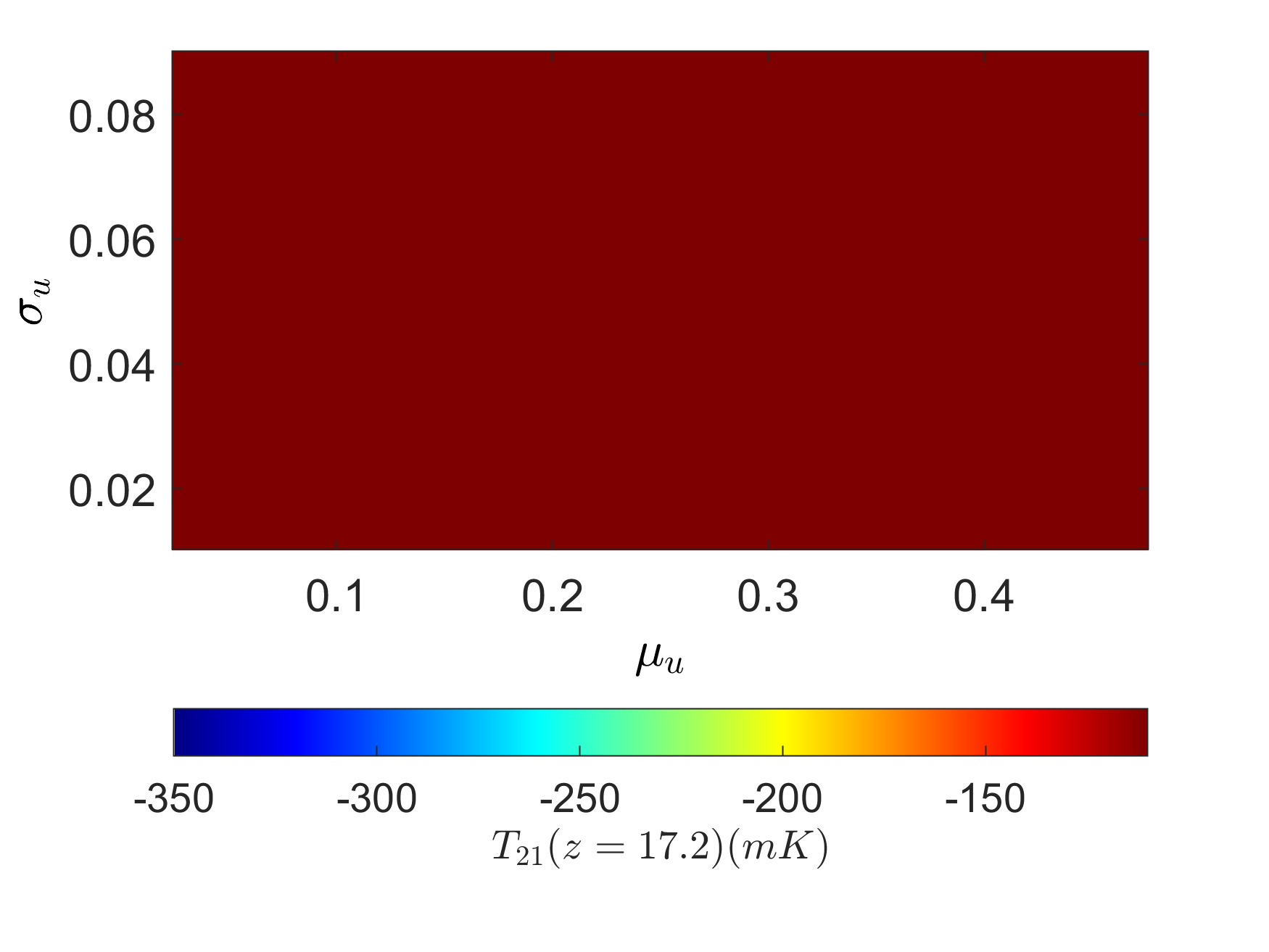}\\		
    \end{tabular}
    \caption{\label{fig:T21_cont_u} Contour representation of $T_{21} (z=17.2)$ (mK) in $\sigma_u - \mu_u$ plane for (a) $\mu_o=0.51$, $\sigma_o=0.01$ (b) $\mu_o=0.51$, $\sigma_o=0.09$, (c) $\mu_o=0.99$, $\sigma_o=0.01$, (d) $\mu_o=0.99$, $\sigma_o=0.09$.}
\end{figure*}

In (\autoref{fig:T21_cont_u}), the variation of $T_{21}$ at $z = 17.2$ in the $\mu_u-\sigma_u$ plane is shown for different sets of values of $(\mu_o,\sigma_o)$ using a contour plot. The value of $\mu_u$ varies in the range of $0 - 0.5$ along the x-axis while $\sigma_u$ is varied along the y-axis in the range of $0.01 - 0.09$. In subplots (a) and (b) of the figure, $\mu_o$ is fixed at $0.51$, and $\sigma_o$ has the values of $0.01$ and $0.09$, respectively. In subplots (c) and (d) of the figure, $\mu_o$ is fixed at $0.99$ while $\sigma_u$ has the values of $0.01$ and $0.09$, respectively. The effect of varying $\sigma_o$ while keeping $\mu_o$ fixed can also be seen from these two subplots. From subplots (a) and (c), it can be seen that changing the value of $\mu_o$ keeping $\sigma_o$ fixed has a slightly more prominent effect on the brightness temperature than vice versa. The lowest value of $T_{21}(z=17.2)$ of around $-320$ mK is obtained in the bottom left part of the subplot (a) of the figure for $\mu_u \approx 0.01$. This also affirms our analysis of (\autoref{fig:T21}) that lower values of $\sigma_u$, $\sigma_o$, $\mu_o$ and $\mu_u$ lead to lower values (more negative) of $T_{21}(z=17.2)$. This figure also highlights the insignificance of overdense parameters as all four subplots are very much alike, so changing the overdense parameters does not affect the output much.

\section{Observational Constraints}\label{sec:obs}

We now examine the backreaction framework in the context of our model
of multiple subregions with respect to observational data and determine the optimum values of our model parameters. We perform a Bayesian analysis to compare our model with the Union 2.1 supernova Ia distance modulus versus redshift data \cite{union}. To compare our model with the observational data, we employ the earlier-mentioned covariant scheme (\autoref{eq:covariant_sch_1}, \autoref{eq:covariant_sch_2}). The first equation of the covariant scheme (\autoref{eq:covariant_sch_1}) relates the theoretically calculated quantity from our model $a_{\mathcal{D}}$ with the cosmological redshift, $z$, and the second equation (\autoref{eq:covariant_sch_2}) relates the theoretically calculated quantity from our model $\langle\rho\rangle_{\mathcal{D}}$ with the observational quantity, the angular diameter distance $D_A$. From this $D_A$, we can calculate the distance modulus using standard cosmological distance relations and thus compare our model with the Union 2.1 supernova Ia data.

\begin{figure*}
    \centering
    \includegraphics[width=\textwidth]{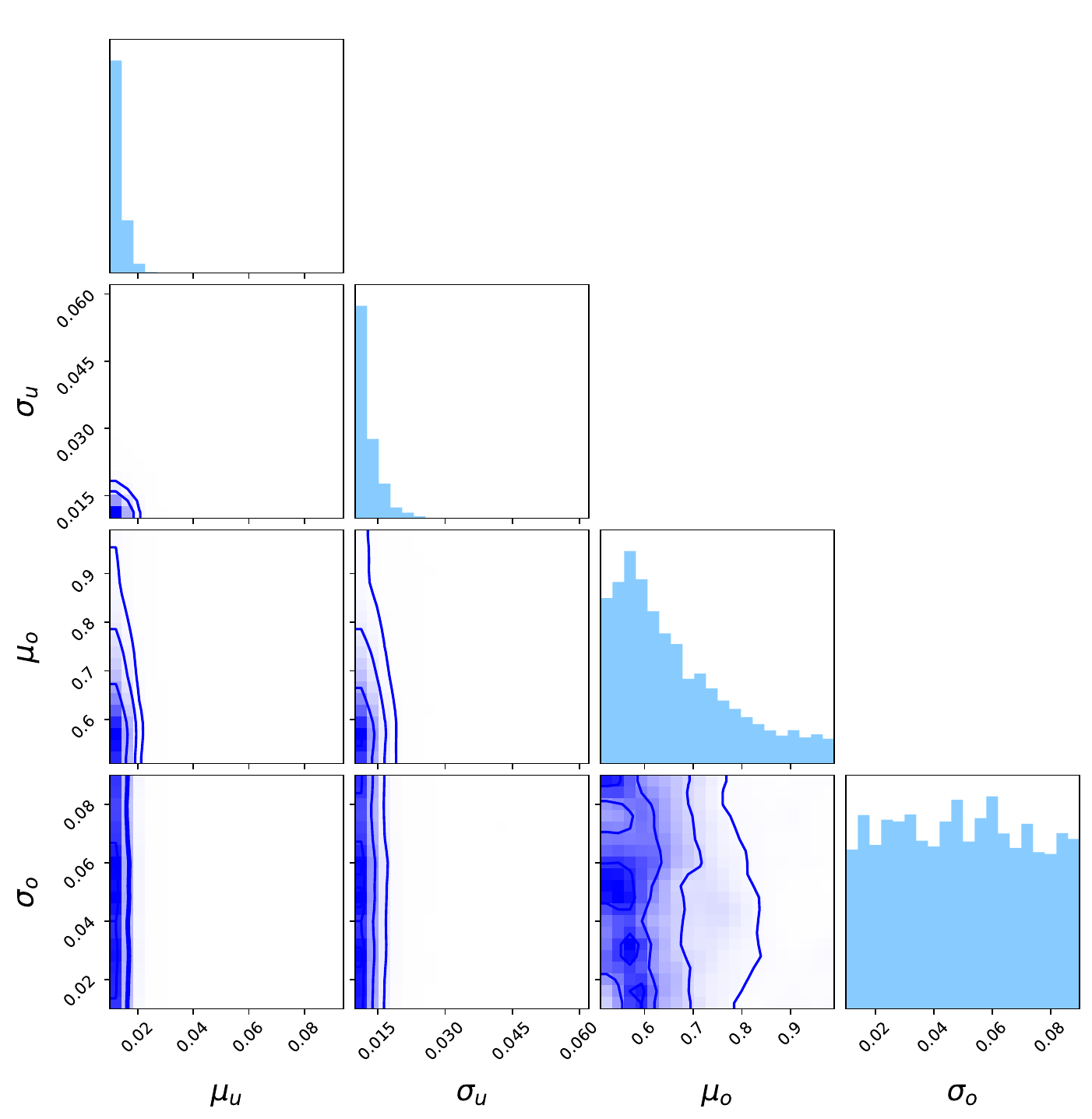}
    \caption{\label{fig:union} Corner plot showing the MCMC result for our model carried out using the observational results of the Union 2.1 supernova Ia data \cite{union}. The diagonal histograms show the marginalized posterior densities for each parameter. }
\end{figure*}

In this analysis, the resulting posterior distributions of different parameters are obtained by the Markov Chain Monte Carlo (MCMC) iteration method (\autoref{fig:union}) by using the \texttt{MCMCSTAT} package \cite{mcmc1,mcmc2}. We use a total of $3\times10^3$ number of events with the adaptation interval of $100$, within the parameter range: $\mu_u \in [0.01,0.49]$, $\sigma_u \in [0.01, 0.09]$, $\mu_o \in [0.51, 0.99]$ and $\sigma_o \in [0.01, 0.09]$. The topmost plots of the first, second, third and fourth columns of (\autoref{fig:union}) represent the posterior distribution of the parameters $\mu_u$, $\sigma_u$, $\mu_o$ and $\sigma_o$, respectively, obtained by marginalizing the other parameters. The other plots of (\autoref{fig:union}) show the contour representation of the posterior distribution in different sets of a two-parameter space. In these contour plots, the darker-coloured regions denote higher posterior regions, and the lines indicate the boundaries of $1\sigma$, $2\sigma$ and $3\sigma$ regions, respectively. The diagonal panels show the 1-D histogram of the posterior distribution for each model parameter obtained by marginalizing the other parameters. The off-diagonal panels show 2-D projections of the posterior probability distributions for each pair of parameters and correlations between the parameters and contours.

 From this analysis the obtained set of optimum points are $\mu_u = 0.01^{+0.00}_{-0.00}$, $\sigma_u = 0.01^{+0.00}_{-0.00}$, $\mu_o = 0.63^{+0.16}_{-0.08}$ and $\sigma_o = 0.05^{+0.03}_{-0.03}$ respectively. These optimum points are obtained considering all four parameters. However, from the marginalized posterior plot for $\mu_o$ (in (\autoref{fig:union})), one can notice that the most probable value of $\mu_o$ is slightly lower ($\mu_o=0.55$) than the corresponding optimal point. This is because posterior plots for each parameter plotted along the diagonal are obtained by marginalizing the other parameters. These plots do not consider other parameters; therefore, the most probable values differ from the optimum values. Similarly, the most probable values for other model parameters are $\mu_u = 0.01$, $\sigma_u = 0.01$ and $\sigma_o = 0.06$. It can be seen that lower values of $\sigma_u$, $\mu_u$ and $\mu_o$ are favored, which in turn favors a reduced brightness temperature of the $T_{21}$ signal, as determined from our analysis of (\autoref{fig:T21} - \autoref{fig:T21_cont_u}).

\begin{figure}
    \centering
    \includegraphics[width=0.5\textwidth]{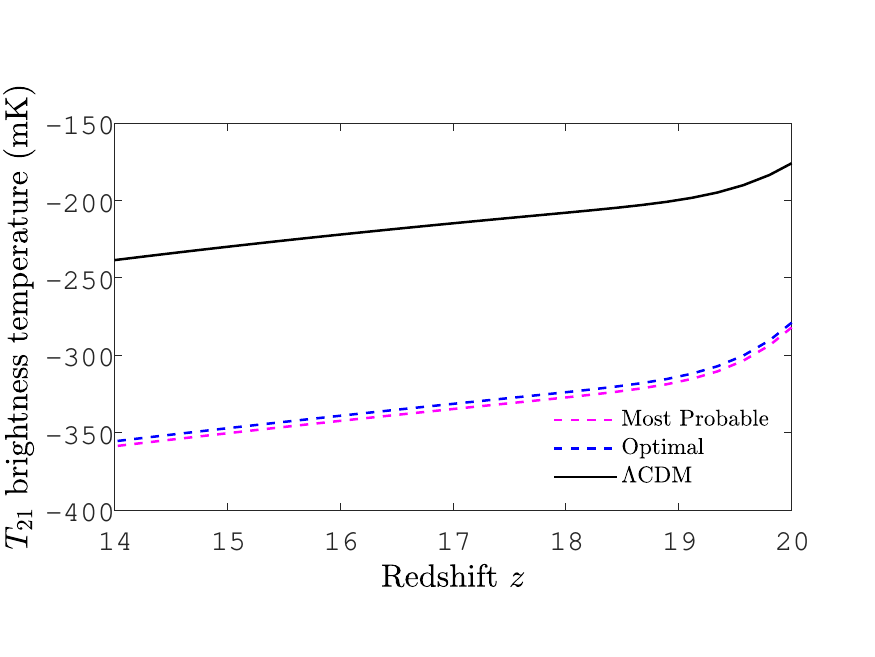}
    \caption{\label{fig:optimal_probable} Plots of brightness temperature $T_{21}$ (mK) for the $\Lambda$CDM model and our backreaction model for the optimal values and the most probable values of our model parameters obtained from the MCMC analysis for the redshift range 14-20. Dashed plotlines are for our backreaction model.}
\end{figure}

In (\autoref{fig:optimal_probable}), the brightness temperature $T_{21}$ is plotted in units of mK as a function of redshift $z$ in the range $14 - 20$   for the $\Lambda$CDM model and for the optimal and most probable values of the parameters of our backreaction model, obtained from the MCMC analysis. The set of optimal values used is $(\mu_u,\sigma_u,\mu_o,\sigma_o) = (0.01,0.01,0.63,0.05)$ and the set of most probable value used is $(\mu_u,\sigma_u,\mu_o,\sigma_o) = (0.01,0.01,0.55,0.06)$. Both the most probable and optimal sets of values gives a brightness temperature considerably lower than the $\Lambda$CDM model for this redshift range. At $z = 14$, $T_{21}$ for our backreaction model, the optimal and most probable set of parameter values is around $\approx-350mK$ which is within the range of the EDGES result. This value is lower than the $\approx-240$mK given by the $\Lambda$CDM model.

\section{Conclusions}\label{sec:results}

Recent observations indicate that our Universe contains an inhomogeneous matter distribution at considerably large scales \cite{Labini_2009, wiegand_scale, lopez}. The effect of these inhomogeneities on various cosmological phenomena calls for close scrutiny. In the present study, we revisit the 21-cm cosmology \cite{FURLANETTO2006181, morales_review, Pritchard_2012}, in a spacetime with matter distribution inhomogeneities. We explore the brightness temperature of the 21-cm signal as a function of the redshift under the impact of backreaction from matter inhomogeneities.

In our analysis, we use the widely used Buchert formalism \cite{Buchert, Buchert2001} of averaging over inhomogeneities to evaluate the backreaction effect. The Buchert framework facilitates the relation of theoretically evaluated quantities with observables such as redshift and angular diameter distance \cite{rasanen1, rasanen2, Koksbang_2019, Koksbang2, Koksbang3}. Within this
framework, we construct a model of multiple subregions with a Gaussian distribution of parameters to mimic the actual Universe containing multiple voids and structures at the present epoch. We employ the covariant scheme to relate our theoretically evaluated parameters with observational quantities. 

Using this model, we calculate the brightness temperature $T_{21}$ of the 21-cm signal as a function of the redshift and analyze it for our model parameters. Such a model of spacetime that we have employed leads to a modification of the Hubble evolution, making it desirable to constrain our model parameters using observational results. To correlate our model with observation data, we obtain the marginalized posterior densities for each model parameter through Markov Chain Monte Carlo (MCMC) simulations using Union 2.1 supernova Ia data \cite{union}. 

Our analysis shows that the 21-cm brightness temperature $T_{21}$ could be lowered by a significant amount under the impact of altered Hubble evolution resulting due to backreaction from matter distribution inhomogeneities. Such a result follows without utilizing any exotic physics or non-standard models of dark matter and dark energy, such as schemes employed in a host of previous works \cite{Ashadul_mnras, Ashadul_PRD, Halder_2021, Clark_2018, Datta_PRD, yang2020, munoz_haimound, Halder_2021, Datta_PRD, Halder_2022} to lower the 21-cm brightness temperature. In particular, using the optimal and most probable values of our model parameters obtained through the MCMC simulations, it can be seen that $T_{21}$ could drop to levels below the predictions of $\Lambda$CDM and within the range of the EDGES result.

We conclude by noting that several earth-based and space-based experiments have been proposed to observe and record the 21-cm signal effectively \cite{Burns_2017, DARE, Pratush, REACH, BIGHORNS}. Our analysis obtains the evolution of $T_{21}$ over a wide range of redshift $z$ in (\autoref{fig:T21}). Thus, it can be used in conjugation with data from such experiments to analyze the role of matter distribution inhomogeneities in 21-cm cosmology. Specifically, if any dip of temperature below the $\Lambda$CDM prediction is observed in the 21-cm signal, our present analysis should motivate further detailed investigations of other backreaction scenarios \cite{Ellis1984, Futamase, Zalaletdinov1992, Zalaletdinov1993, Gasperini_2011}, as well.

\section{Acknowledgements}
The authors would like to thank the anonymous referee for insightful comments. SSP would like to thank the Council of Scientific and Industrial Research (CSIR), Govt. of India, for funding through the CSIR-SRF-NET fellowship.

\appendix

\section{Calculation of $t_0$} \label{app:t0_calc}
\par Using (\autoref{eq:averaging}), we can break down $H_D$ as,

\begin{equation}\label{eq:app_HD_break}
    H_\mathcal{D} = \sum_i\lambda_{o_i}H_{o_i} + \sum_j\lambda_{u_j}H_{u_j} = \lambda_oH_o + \lambda_uH_u,
\end{equation}
where, $\lambda_oH_o = \sum_i\lambda_{o_i}H_{o_i}$ represents the collective contribution of overdense subregions, and similarly $ \lambda_uH_u= \sum_i\lambda_{u_i}H_{u_i}$ represents the collective contribution of underdense subregions. Also, using (\autoref{eq:aD3_sum}), we can write, 
\begin{equation}
    a_\mathcal{D}^3 = \lambda_{o,0}a_o^3+\lambda_{u,0}a_u^3
\end{equation}
where, $\lambda_{o,0}a_o^3$ represents the collective contribution of the overdense subregions and similarly $\lambda_{u,0}a_u^3$ for the underdense subregions. Using (\autoref{eq:lambda_o_relation}), (\autoref{eq:app_HD_break}) can be written as
\begin{equation}
\begin{split}
    H_\mathcal{D} &= \lambda_{{o},0}\frac{a^3_{o}}{a^3_\mathcal{D}}H_{o} + \lambda_{{u},0}\frac{a^3_{u}}{a^3_\mathcal{D}}H_{u}\\
    &= H_o\left(\frac{\lambda_{{o},0}a^3_{o}}{\lambda_{o,0}a_o^3+\lambda_{u,0}a_u^3} + \frac{\lambda_{{u},0}a^3_{u}}{\lambda_{o,0}a_o^3+\lambda_{u,0}a_u^3}\frac{H_{u}}{H_o}\right)\\
    &= H_o(1-v+vh),
\end{split}
\end{equation}
where, using the definitions of $h$ and $v$ from \cite{Koksbang2023}, we have defined similarly $h := H_u/H_o$ and $v := \frac{\lambda_{{u},0}a^3_{u}}{\lambda_{o,0}a_o^3+\lambda_{u,0}a_u^3}$. Our definition of $v$ is different from theirs due to the different scaling of $a_\mathcal{D}$ that we have used here. 
Therefore,
\begin{equation}
    H_\mathcal{D} =  H_o(1-v+vh) = \frac{\sum_i\lambda_{o_i}H_{o_i}}{\lambda_o}(1-v+vh)
\end{equation}

Now, using (\autoref{eq:ao}) and (\autoref{eq:ao_t}),
\begin{equation}
    H_{o_i} = \frac{\sin{\phi_{o_i}}(\phi_{o_i} - \sin{\phi_{o_i}})}{t_0(1 - \cos{\phi_{o_i}})^2}
\end{equation}

Therefore,
\begin{equation}
    H_\mathcal{D} = \frac{1}{\lambda_o t_0} (1 - v + vh)\sum_i \lambda_{o_i}\frac{\sin{\phi_{o_i}}(\phi_{o_i} - \sin{\phi_{o_i}})}{(1 - \cos{\phi_{o_i}})^2}
\end{equation}
At present time, $t_0$,
\begin{equation}
     H_{\mathcal{D}_0} = \frac{1}{\lambda_{o,0} t_0} (1 - v_0 + v_0 h_0)\sum_i \lambda_{o_{i,0}}\frac{\sin{\phi_{o_{i,0}}}(\phi_{o_{i,0}} - \sin{\phi_{o_{i,0}}})}{(1 - \cos{\phi_{o_{i,0}}})^2}
\end{equation}

We see that,
\begin{equation}\label{eq:app_t0_final}
    t_0 =\frac{1}{\lambda_{o,0} H_{\mathcal{D}_0}} (1 - v_0 + v_0 h_0)\sum_i \lambda_{o_{i,0}}\frac{\sin{\phi_{o_{i,0}}}(\phi_{o_{i,0}} - \sin{\phi_{o_{i,0}}})}{(1 - \cos{\phi_{o_{i,0}}})^2}
\end{equation}
so we need to fix either $t_0$ or $H_{\mathcal{D}_0}$. Here, we have chosen $H_{\mathcal{D}_0}$ = 70 km/s/Mpc.
Now, we have,
\begin{equation*}
    v_0 = \frac{\lambda_{{u},0}a^3_{u,0}}{\lambda_{o,0}a_{o,0}^3+\lambda_{u,0}a_{u,0}^3}; \hspace{1 cm} h_0 = \frac{H_{u,0}}{H_{o,0}}
\end{equation*}

We have defined our model in such a way that, $a_{o,0} = a_{u,0} = a_{\mathcal{D},0} = 1$. Also, $(\lambda_{o,0},\lambda_{u,0}) = (0.09,0.91)$ \cite{Weigand_et_al}. Therefore, $v_0 = 0.91$. Also,
\begin{widetext}
    \begin{equation}\label{eq:t0_long_o}
    \begin{split}
    \lambda_oH_o = \sum_i\lambda_{o_i}H_{o_i} 
    \implies H_o = \frac{1}{\lambda_o}\sum_i\lambda_{o_i}H_{o_i}\\
    \implies H_{o,0} = \frac{1}{\lambda_{o,0}}\sum_i\lambda_{{o_i},0}H_{o_{i,0}}
    = \frac{1}{\lambda_{o,0}\times t_0}\sum_i\left(\dfrac{N_o}{\sigma_o \sqrt{2 \pi}} e^{-(q_{o_{i,0}}-\mu_o)^2/2\sigma_o^2}\times\frac{\sin{\phi_{o_{i,0}}}(\phi_{o_{i,0}} - \sin{\phi_{o_{i,0}}})}{(1 - \cos{\phi_{o_{i,0}}})^2}\right)
    \end{split}
    \end{equation}
\end{widetext}
where we have used (\autoref{eq:gauss_o}). Similarly, using (\autoref{eq:au}), (\autoref{eq:au_t}) and (\autoref{eq:gauss_u}), we get the following,
\begin{widetext}
    \begin{equation}\label{eq:t0_long_u}
    \begin{split}
    H_{u,0} = \frac{1}{\lambda_{u,0}\times t_0}
    \sum_i\left(\dfrac{N_u}{\sigma_u \sqrt{2 \pi}} e^{-(q_{u_{i,0}}-\mu_u)^2/2\sigma_u^2}
    \times\frac{\sinh{\phi_{u_{i,0}}}(\sinh{\phi_{u_{i,0}}} - \phi_{u_{i,0}})}{(\cosh{\phi_{u_{i,0}}} - 1)^2}\right)
    \end{split}
    \end{equation} 
\end{widetext}
The $t_0$ term in the denominators of (\autoref{eq:t0_long_o}) and (\autoref{eq:t0_long_u}) cancels out and all the other quantities are known. Therefore, we have,
\begin{widetext}
    \begin{equation}\label{eq:app_h0_final}
    h_0 = \frac{H_{u,0}}{H_{o,0}} = \frac{\lambda_{o,0}}{\lambda_{u,0}}\frac{\sum_i\left(\dfrac{N_u}{\sigma_u \sqrt{2 \pi}} e^{-(q_{u_{i,0}}-\mu_u)^2/2\sigma_u^2}\times\frac{\sinh{\phi_{u_{i,0}}}(\sinh{\phi_{u_{i,0}}} - \phi_{u_{i,0}})}{(\cosh{\phi_{u_{i,0}}} - 1)^2}\right)}{\sum_i\left(\dfrac{N_o}{\sigma_o \sqrt{2 \pi}} e^{-(q_{o_{i,0}}-\mu_o)^2/2\sigma_o^2}\times\frac{\sin{\phi_{o_{i,0}}}(\phi_{o_{i,0}} - \sin{\phi_{o_{i,0}}})}{(1 - \cos{\phi_{o_{i,0}}})^2}\right)}
    \end{equation}
\end{widetext}

So, from (\autoref{eq:app_h0_final}), we can calculate $h_0$ and then using this value of $h_0$ and $v_0$ in (\autoref{eq:app_t0_final}), we can obtain $t_0$. In this way, using the above procedure, we can fix $t_0$ for a given set of values of our model parameters $(\mu_u,\sigma_u,\mu_o,\sigma_o)$.

\section{Our multidomain model} \label{app:model}

Our multidomain model is made up of multiple subdomains of overdense and underdense. The underdense subdomains are negatively curved FLRW regions while the overdense subdomains are positively curved FLRW regions having densities greater than those of the underdense subdomains. Both of these subregions are composed of dust. Now, the scale factor for these overdense and underdense subdomains as functions of cosmic time $t$ is given by (\autoref{eq:ao}, \autoref{eq:ao_t}) and (\autoref{eq:au}, \autoref{eq:au_t}) respectively.

The densities of underdense subregions is given by \cite{weinberg},
\begin{equation}
    \rho_{u_i} = \frac{\rho_{u_{i,0}}}{a_{u_i}^3} \label{eq:rho_underdense}
\end{equation}
where $\rho_{u_{i,0}}$ is the density at present time $t_0$, which is given by,
\begin{equation}
    \frac{\rho_{u_{i,0}}}{\rho_{u_{i,c}}} = q_{u_{i,0}}; \hspace{0.5 cm} \mbox{where} \hspace{0.5 cm} \rho_{u_{i,c}} = \frac{3H_{u_i,0}^2}{8\pi G}
    \label{eq:rho0_underdense}
\end{equation}
where $\rho_{u_{i,c}}$ is the critical density and $H_{u_i,0}$ is the value of present time Hubble parameter for the $i^{th}$ underdense subregion.

Densities for the overdense subregions can also be defined similarly,
\begin{equation}
    \rho_{o_i} = \frac{\rho_{o_{i,0}}}{a_{o_i}^3} \label{eq:rho_overdense}
\end{equation}
where $\rho_{o_{i,0}}$ is the density at present time $t_0$, which is given by,
\begin{equation}
    \frac{\rho_{o_{i,0}}}{\rho_{o_{i,c}}} = q_{o_{i,0}}; \hspace{0.5 cm} \mbox{where} \hspace{0.5 cm} \rho_{o_{i,c}} = \frac{3H_{o_i,0}^2}{8\pi G}
    \label{eq:rho0_overdense}
\end{equation}
where $\rho_{o_{i,c}}$ is the critical density and $H_{o_i,0}$ is the value of present time Hubble parameter for the $i^{th}$ overdense subregion.

Choosing our parameters in the range $0 < q_{u_{i,0}} < 0.5$ and
$1/2 < q_{o_{i,0}} < 1$, it follows from \autoref{eq:rho0_underdense} and
\autoref{eq:rho0_overdense} that the densities of underdense subdomains remain always less than the densities of overdense subdomains.

The combined density for all overdense subregions ($\rho_o$) is given by,
\begin{equation}
    \lambda_o\rho_o = \sum_i\lambda_{o_i}\rho_{o_i} \implies \rho_o = \frac{\sum_i\lambda_{o_i}\rho_{o_i}}{\sum_i\lambda_{o_i}} \label{eq:rho_o}
\end{equation}
where $\lambda_{o_i}$ are defined in (\autoref{eq:lambda_o_relation}) and $\lambda_o$ is the total volume fraction of the overdense subregions. Similarly,
\begin{equation}
    \lambda_u\rho_u = \sum_i\lambda_{u_i}\rho_{u_i} \implies \rho_u = \frac{\sum_i\lambda_{u_i}\rho_{u_i}}{\sum_i\lambda_{u_i}} \label{eq:rho_u}
\end{equation}
where $\lambda_{u_i}$ are defined in (\autoref{eq:lambda_u_i}) and $\lambda_u$ is the total volume fraction of the underdense subregions.
\par Also, $\langle\rho\rangle_\mathcal{D}$, the averaged density of the global domain is given by
\begin{equation}
    \begin{split}
    \langle\rho\rangle_\mathcal{D} & = \lambda_o\rho_o + \lambda_u\rho_u\\
    & = \sum_i \lambda_{o_i}\rho_{o_i} + \sum_i\lambda_{u_i}\rho_{u_i} \label{eq:rho_D}
\end{split}
\end{equation}

The various density parameters are obtained in our analysis in the following way. From (\autoref{eq:aD_dot}), we have
\begin{equation}
    3H^2_{\mathcal{D}} = 8\pi G\langle\rho\rangle_{\mathcal{D}} - \frac{1}{2}\langle\mathcal{R}\rangle_{\mathcal{D}} - \frac{1}{2}\mathcal{Q}_{\mathcal{D}} \label{eq:app_aD_dot}
\end{equation}
Dividing by $3H^2_{\mathcal{D}}$ throughout, we get,
\begin{equation}\label{eq:app_sum_omega}
    1 = \Omega^{\mathcal{D}}_m + \Omega^\mathcal{D}_\mathcal{R} + \Omega^\mathcal{D}_\mathcal{Q}
\end{equation}
where,
\begin{equation}\label{eq:app_omega}
    \Omega^\mathcal{D}_m = \frac{8\pi G\langle\rho\rangle_\mathcal{D}}{3H^2_{\mathcal{D}}}; \hspace{0.5 cm}\Omega^\mathcal{D}_\mathcal{R} = -\frac{\langle\mathcal{R}\rangle_\mathcal{D}}{6H_\mathcal{D}^2};\hspace{0.5 cm} \Omega^\mathcal{D}_\mathcal{Q} = -\frac{\mathcal{Q}_\mathcal{D}}{6H_\mathcal{D}^2}
\end{equation}

From (\autoref{eq:QDsum2}), we can calculate $\mathcal{Q}_\mathcal{D}$ for our model and then from (\autoref{eq:app_omega}), $\Omega^\mathcal{D}_\mathcal{Q}$ can be calculated. Then using (\autoref{eq:rho_D}) and (\autoref{eq:app_omega}), $\Omega^\mathcal{D}_m$ can be calculated and $\Omega^\mathcal{D}_\mathcal{R}$ can be calculated from (\autoref{eq:app_sum_omega}).
In this way, all the average density parameters can be obtained.

\bibliography{21cm_inhomo1.bib}

\end{document}